\RequirePackage{fix-cm}
\documentclass[smallcondensed]{svjour3}     %
\smartqed  %
\usepackage{graphicx}
\usepackage{subfig}
\usepackage{caption}
\usepackage{multirow}
\usepackage{gensymb}
\usepackage{dblfloatfix}
\usepackage{bm}
\usepackage{booktabs}
\usepackage{url}
\usepackage{lineno}
\usepackage{adjustbox}
\usepackage[shortlabels]{enumitem}
\usepackage{orcidlink}
\newcommand{\expnumber}[2]{{#1}\mathrm{e}{#2}}
\journalname{Autonomous Agents and Multi-Agent Systems}
\begin{document}

\title{Warmth and competence in human-agent cooperation\thanks{An earlier version of this work was presented at the 2022 International Conference on Autonomous Agents and Multi-Agent Systems (AAMAS 2022), and can be found online at \protect\url{https://arxiv.org/abs/2201.13448/v2}.}
}

\author{Kevin R. McKee~\orcidlink{0000-0002-4412-1686} \and
        Xuechunzi Bai~\orcidlink{0000-0002-2277-5451}         \and
        Susan~T.~Fiske~\orcidlink{0000-0002-1693-3425}
}

\authorrunning{Kevin R. McKee, Xuechunzi Bai, \& Susan T. Fiske} %

\institute{ K. McKee \at
            \email{kevinrmckee@google.com}
}

\date{Received: 15 Feb 2023 / Accepted: 12 Apr 2024}

\maketitle

\begin{abstract}
Interaction and cooperation with humans are overarching aspirations of artificial intelligence (AI) research.
Recent studies demonstrate that AI agents trained with deep reinforcement learning are capable of collaborating with humans.
These studies primarily evaluate human compatibility through ``objective'' metrics such as task performance, obscuring potential variation in the levels of trust and subjective preference that different agents garner.
To better understand the factors shaping subjective preferences in human-agent cooperation, we train deep reinforcement learning agents in Coins, a two-player social dilemma.
We recruit $N = 501$ participants for a human-agent cooperation study and measure their impressions of the agents they encounter.
Participants' perceptions of warmth and competence predict their stated preferences for different agents, above and beyond objective performance metrics.
Drawing inspiration from social science and biology research, we subsequently implement a new ``partner choice'' framework to elicit \textit{revealed} preferences: after playing an episode with an agent, participants are asked whether they would like to play the next episode with the same agent or to play alone.
As with stated preferences, social perception better predicts participants' revealed preferences than does objective performance.
Given these results, we recommend human-agent interaction researchers routinely incorporate the measurement of social perception and subjective preferences into their studies.
\keywords{Human-agent cooperation \and Human-agent interaction \and Warmth \and Competence \and Social perception \and Partner choice \and Preferences}
\end{abstract}

\section{Introduction}

Trust is central to the development and deployment of artificial intelligence (AI)~\cite{jobin2019global,stanton2021trust}. However, many members of the public harbor doubts and concerns about the trustworthiness of AI~\cite{cave2019hopes,dietvorst2015algorithm,fast2017long,kelley2021happy}.
This presents a pressing issue for cooperation between humans and AI agents~\cite{dafoe2020open}.

Algorithmic development research has been slow to recognize the importance of trust and preferences for cooperative agents. Recent studies show that deep reinforcement learning can be used to train interactive agents for human-agent collaboration~\cite{carroll2019utility,lockhart2020human,strouse2021collaborating,tylkin2021learning}. The ``human-compatible'' agents from these experiments demonstrate compelling improvements in game score, task accuracy, and win rate over established benchmarks. However, a narrow focus on ``objective'' metrics of performance obscures any differences in subjective preferences humans develop over cooperative agents. Two agents may generate similar benefits in terms of typical performance metrics, but human teammates may nonetheless express a strong preference for one over the other~\cite{siu2021evaluation,strouse2021collaborating}. Developing human-compatible, cooperative agents will require evaluating agents on dimensions other than objective performance.

\begin{figure}[!t]
    \adjustbox{valign=b}{%
    \begin{minipage}[t]{.55\linewidth}
    	\centering
    	\subfloat[New agent encounter. \label{fig:new_situation}]{\includegraphics[width=0.95\textwidth]{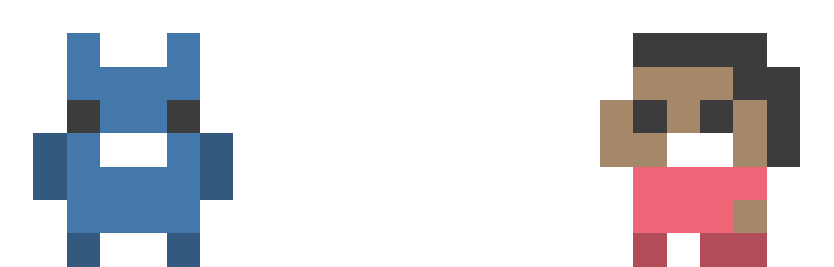}} \\
    	\subfloat[Social perception. \label{fig:social_perception}]{\includegraphics[width=0.95\textwidth]{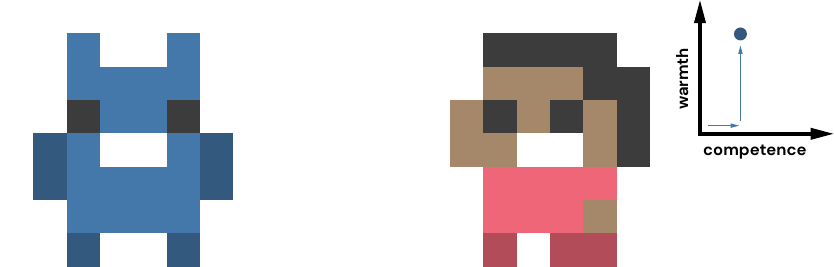}} \\
        \subfloat[Partner choice. \label{fig:partner_choice}]{\includegraphics[width=0.95\textwidth]{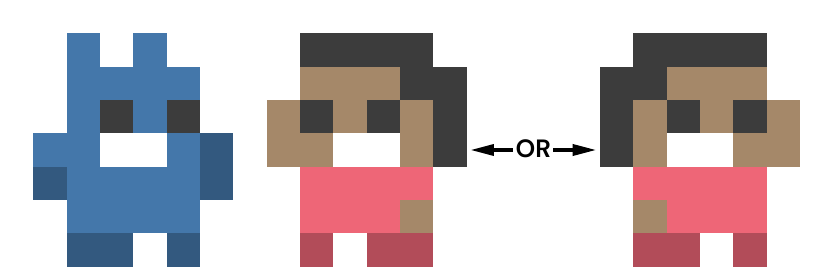}}
    \end{minipage}}
    \hfill
    \adjustbox{valign=b}{%
    \begin{minipage}[t]{.4\linewidth}
    	\caption{When humans encounter a new agent, they automatically and rapidly form an impression of the agent (social perception). The human can leverage their impression to decide whether to continue or discontinue the interaction (partner choice). \label{fig:intro}}
	\end{minipage}}
	\hspace{1em}
\end{figure}

What shapes subjective preferences for artificial agents, if not a direct mapping of agent performance? One possible source of variance is \textit{social perception}. When encountering a novel actor, humans rapidly and automatically evaluate the actor along two underlying dimensions: warmth and competence~\cite{abele2021navigating,fiske2007universal,fiske2018stereotype,mckee2023humans}. These perceptions help individuals ``make sense of [other actors] in order to guide their own actions and interactions''~\cite{fiske1993social} (Figure~\ref{fig:intro}). The competence dimension aligns with the established focus on performance and score in machine learning research~\cite{birhane2022values}: How effectively can this actor achieve its interests? Appraising an actor's warmth, on the other hand, raises a novel set of considerations: How aligned are this actor's goals and interests with one's own? Research on social cognition consistently demonstrates that humans prefer others who are not only competent, but also warm~\cite{abele2021navigating,fiske2002model}. Hence, we predict that perceived warmth will be an important determinant of preferences for artificial agents.

Here we run behavioral experiments to investigate social perception and subjective preferences in human-agent interaction. We train reinforcement learning agents to play Coins, a mixed-motive game, varying agent hyperparameters known to influence cooperative behavior and performance in social dilemmas. Three co-play experiments then recruit human participants to interact with the agents, measure participants' judgments of agent warmth and competence, and elicit participant preferences over the agents.

Experiments evaluating human views of agents frequently rely on stated preferences, typically by directly asking participants which of two agents they preferred as a partner~\cite{du2020ave,siu2021evaluation,strouse2021collaborating}. Such self-report methods can be insightful tools for research~\cite{paulhus2007self}. However, they exhibit limited ecological validity and are vulnerable to experimenter demand~\cite{de2018measuring}. In this paper, we overcome these challenges by eliciting \textit{revealed preferences}~\cite{samuelson1938note}: Do people even want to interact with a given agent, if given the choice not to? Partner choice, or the ability to leave or reject an interaction, is a well-established revealed-preference paradigm in evolutionary biology and behavioral economics~\cite{barclay2007partner,baumard2013mutualistic,brown2004relational,slonim2008increases}. {While studies that measure revealed preferences (e.g.,~\cite{kox2021trust,ramchurn2016human}) are not inherently immune to experimenter demand, partner-choice measures can mitigate demand effects when participants are compensated based on their performance (``incentivized choice''; see~\cite{de2018measuring}).} Partner choice also carries external validity for interaction research: in the context of algorithmic development, we can view partner choice as a stand-in for the choice to adopt an artificial intelligence system~\cite{beaudry2010other,davis1989perceived,parasuraman1997humans}. {Users may test out several suggestions from a recommender system before deciding whether or not to rely on it for future decisions. Similarly, commuters might tentatively try several rides from self-driving cars to help choose whether to transition away from traditional driving.} Overall, partner-choice study designs empower participants with an ability to embrace or leave an interaction with an agent---and thus incorporate an ethic of autonomy~\cite{berlin1969four,jobin2019global,miller1983constraints} into human-agent interaction research.

In summary, this paper makes the following contributions to cooperative AI research:
\begin{enumerate}
    \item {Demonstrates the use of reinforcement learning to train human-compatible agents for a temporally and spatially extended mixed-motive game.}
    \item {Connects human-AI interaction research to frameworks from psychology and economics, identifying tools that researchers can easily import for their own studies.}
    \item Illustrates how social perceptions affect stated and revealed preferences in incentivized interactions with agents, above and beyond standard performance metrics.
\end{enumerate}

\section{Methods}

\subsection{Task}
\label{sec:task}

\begin{figure}[!t]
	\centering
    \includegraphics[width=0.65\textwidth]{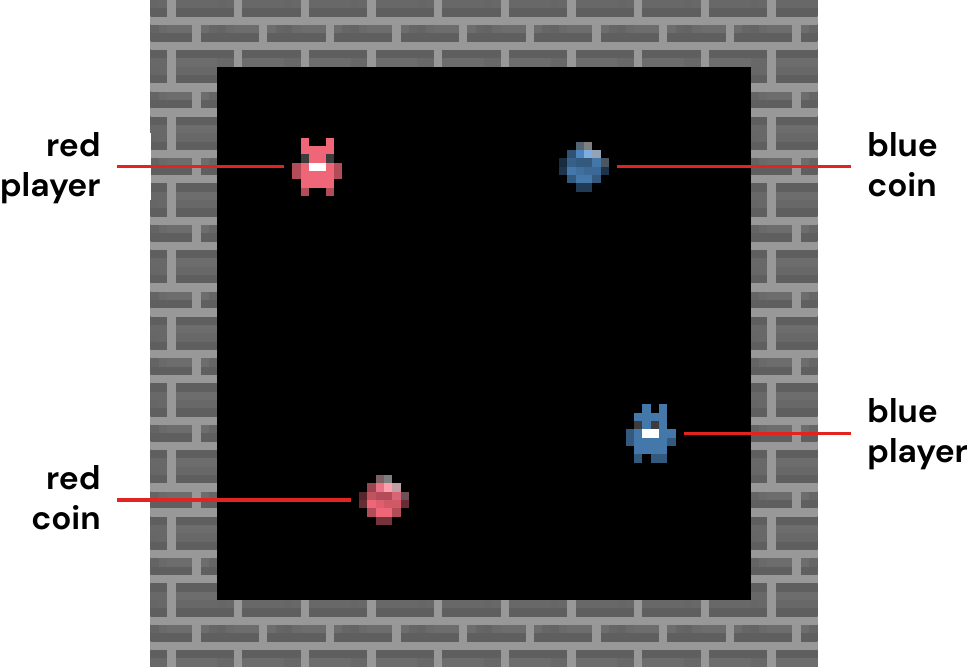}
	\caption{Screenshot of gameplay in Coins. Two players move around a small room and collect colored coins. Coins randomly appear in the room over time. Players receive reward from coin collections depending on the match or mismatch between their color and the coin color.}
	\label{fig:coins}
\end{figure}

Coins~\cite{foerster2018learning,gemp2022d3c,lerer2017maintaining,peysakhovich2018consequentialist} is a mixed-motive Markov game~\cite{littman1994markov} played by $n = 2$ players (Figure~\ref{fig:coins}{; see also Appendix~\ref{sec:app/task} for full task details}).
Players of two different colors occupy a small gridworld room with width $w$ and depth $d$. Coins randomly appear in the room throughout the game, spawning on each cell with probability $P$. Every coin matches one of the two players in color. On each step of the game, players can stand still or move around the room.

The goal of the game is to earn reward by collecting coins. Players pick up coins by stepping onto them. {Each coin collection generates reward for both players} as a function of the match or mismatch between the coin color and the collecting player's color.
Under the canonical rules (Table~\ref{tab:payoff_table_canonical}), a player receives $+1$ reward for picking up a coin of any color. If a player collects a coin of their own color (i.e., a matching coin), the other player is unaffected. However, if a player picks up a coin of the other player's color (i.e., a mismatching coin), the other player receives $-2$ reward. In the short term, it is always individually advantageous to collect an available coin, whether matching or mismatching. However, {this strategy will lower the score of the other player. Players} achieve the socially optimal outcome by collecting only the coins that match their color.

\begin{table}[h]
    \centering
    \begin{tabular}[h]{c c c}
        \toprule
        \multirow{2}{*}{Coin color} & Reward & Reward for \\
         & for self & co-player \\
        \midrule
        matching & $+1$ & $+0$ \\
        mismatching & $+1$ & $-2$ \\
        \bottomrule
    \end{tabular}
    \caption{Canonical incentive structure for Coins.}
    \label{tab:payoff_table_canonical}    
\end{table} \vspace{-2em}

Two properties make Coins an ideal testbed for investigating perceptions of warmth and competence. First, as a consequence of its incentive structure, Coins is a social dilemma~\cite{kollock1998social}: players can pursue selfish goals or prosocial goals. Second, relative to matrix games like the Prisoner's Dilemma, Coins is temporally and spatially extended~\cite{leibo2017multi}: players can employ a variety of strategies to achieve their goals, with differing levels of efficiency. We hypothesize that these two features offer sufficient affordance for an observer to infer other players' intentions and their effectiveness at enacting their intentions~\cite{blakemore2001perception,reeder2009mindreading,zacks2004using}.

Our experiments use a colorblind-friendly palette, with red, blue, yellow, green, and purple players and coins (Figure~\ref{fig:app/coins_entities/a}). During agent training, we procedurally generate rooms with width $w$ and depth $d$ independently sampled from $\mathcal{U}\{10, 15\}$. Coins appear in each cell with probability $P = 0.0005$. Episodes last for $T = 500$ steps. Each episode of training randomly samples colors (without replacement) for agents.

In our human-agent interaction studies, co-play episodes use $w = d = 11$, $P = 0.0005$, and $T = 300$. Player colors are randomized across the five players (one human participant and four agent co-players) at the beginning of each study session, and held constant across all episodes within the session.

In Study 1, humans and agents play Coins with the canonical rules. In Studies 2 and 3, humans and agents play Coins with a slightly altered incentive structure. Each outcome increases by $+2$ reward, making all rewards in the game non-negative (Table~\ref{tab:payoff_table_offset}). Since all rewards are offset by the same amount, this reward scheme preserves the social dilemma structure in Coins.

\begin{table}[h]
    \centering
    \begin{tabular}[h]{c c c}
        \toprule
        \multirow{2}{*}{Coin color} & Reward & Reward for \\
         & for self & co-player \\
        \midrule
        matching & $+3$ & $+2$ \\
        mismatching & $+3$ & $+0$ \\
        \bottomrule
    \end{tabular}
    \caption{Alternative incentive structure for Coins.}
    \label{tab:payoff_table_offset}    
\end{table} \vspace{-4em}

\subsection{Agent design and training}

We leverage deep reinforcement learning to train four agents for our human-agent cooperation studies {(see Appendix~\ref{sec:app/agent} for full agent details)}. Overall, our study design and measurement tools are agnostic to the algorithmic implementation of the agents being evaluated. For this paper, the agents learn using the advantage actor-critic (A2C) algorithm~\cite{mnih2016asynchronous}. The neural network consists of a convolutional module, a fully connected module, an LSTM with contrastive predictive coding~\cite{hochreiter1997long,oord2018representation}, and linear readouts for policy and value. Agents train for $5 \times 10^7$ steps in self-play, with task parameters as described in Section~\ref{sec:task}. We consider two algorithmic modifications to the agents to induce variance in social perception.

First, we build the Social Value Orientation (SVO) component~\cite{mckee2020social}, an algorithmic module inspired by psychological models of human prosocial preferences~\cite{griesinger1973toward,liebrand1988ring,murphy2014social}, into our agents. The SVO component parameterizes each agent with $\theta$, representing a target distribution over their reward and the reward of other agents in their environment. SVO agents are intrinsically motivated~\cite{singh2005intrinsically} to optimize for task rewards that align with their parameterized target $\theta$. For these experiments, we endow agents with the ``individualistic'' value $\theta = 0\degree$ and the ``prosocial'' value $\theta = 45\degree$.

Second, we add a ``trembling hand''~\cite{cushman2009accidental,selten1975reexamination} component to the agents for evaluation and co-play. With probability $\epsilon$, the trembling-hand module replaces each action selected by the agent with a random action. This component induces inefficiency in maximizing value according to an agent's learned policy and value function. For these experiments, we apply the ``steady'' value $\epsilon = 0$ and the ``trembling'' value $\epsilon = 0.5$.

\begin{table}[h]
    \centering
    \begin{tabular}[h]{c c c}
        \toprule
         & $\epsilon = 0$ & $\epsilon = 0.5$ \\
        \midrule
        \multirow{2}{*}{$\theta = 0\degree$} & low warmth, & low warmth, \\
         & high competence & low competence \\
        \multirow{2}{*}{$\theta = 45\degree$} & high warmth, & high warmth, \\
         & high competence & low competence \\
        \bottomrule
    \end{tabular}
    \caption{Predictions for social perception (warmth and competence) as a function of agent hyperparameters (Social Value Orientation ${\theta}$ and trembling hand ${\epsilon}$).}
    \label{tab:agent_hyperparameters}    
\end{table}  %

Table~\ref{tab:agent_hyperparameters} summarizes the hyperparameter values and predicted effects for the four evaluated agents.

\subsection{Study design for human-agent studies}

We recruited participants from Prolific~\cite{peer2021data,peer2017beyond} for all studies (total $N = 501$; 47.6\% female, 48.4\% male, 1.6\% non-binary or trans; $m_{\mathrm{age}} = 33$, $sd_{\mathrm{age}} = 11$). We received informed consent from all participants across the three studies. {In each study, participants earned a base level of compensation for completing the study, with an additional bonus that varied as a function of their score in the task.} Appendix~\ref{sec:app/study} presents full details of our study design, including independent ethical review and study screenshots.

Overall, our studies sought to explore the relationship between social perception and subjective preferences in Coins. Study 1 approached these constructs using the canonical payoff structure for Coins~\cite{lerer2017maintaining} and an established self-report framework for eliciting (stated) preferences~\cite{strouse2021collaborating}. {This study framework leverages a ``within-participants design'', meaning that each participant interacts with all agents in the study. Within-participants designs increase the data points collected per participant and allow for better control of between-individual variance, thus improving statistical power for a given sample size.}

We subsequently sought to understand whether the findings from Study 1 replicate under a partner choice framework. Does social perception exhibit the same predictive power for revealed preferences as it does for stated preferences? {While within-participants designs offer several statistical advantages, they are not ideal for studying partner choices. Exposing participants to multiple potential partners can introduce order effects, where the exact sequence of interactions influences participants' responses. Within-participants designs may also fatigue participants, potentially compromising the quality of their responses as the study progresses. Consequently, participants' partner choices may be progressively less motivated by genuine preference and more influenced by extraneous factors. Thus, to study revealed preferences, we switched to a ``between-participants design'' in which each participant interacted with one randomly selected agent.} Given that humans respond more strongly to losses than to commensurate gains~\cite{kahneman1979prospect}, we made one additional change, testing participants' partner choices under a shifted incentive structure with all non-negative outcomes (Table~\ref{tab:payoff_table_offset}). To isolate the effects of the change from stated to revealed preferences, we approached these changes in two stages. Study 2 used the same stated-preference approach and within-participants design as Study 1, but incorporated the offset incentive structure. Study 3 then elicited revealed preferences in place of stated preferences.

We tested the following hypotheses in our studies:

\begin{enumerate}[start=1,label={\bfseries H\arabic*.}]
    \item Social perception predicts participants' stated and revealed preferences. That is, human participants will prefer to play with agents they perceive as warm and competent.
    
    \item Social perception predicts participants' stated and revealed preferences, above and beyond the scores that participants receive. That is, participants' social perceptions of agents will contribute to their preferences independently of the scores they receive when playing with the agents.
    
    \item Social perception will correlate positively with the sentiment of participants' verbal impressions of the agents. That is, participants' social perceptions of agents will emerge as positive sentiment in participants' verbal descriptions of the agents. %
\end{enumerate}

\subsubsection{Study 1}

\begin{figure*}[!b]
	\centering
    \subfloat[Social perception.]{\includegraphics[width=0.231\textwidth]{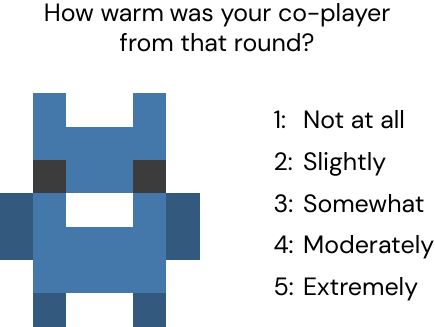} \label{fig:questions/a}} \hfill
    \subfloat[Stated preferences.]{\includegraphics[width=0.411\textwidth]{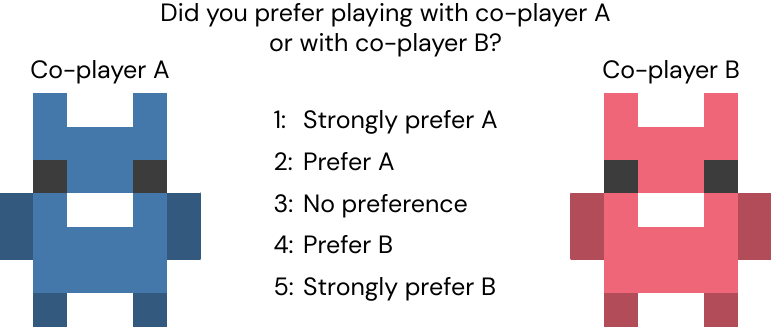} \label{fig:questions/b}} \hfill
    \subfloat[Revealed preferences.]{\includegraphics[width=0.258\textwidth]{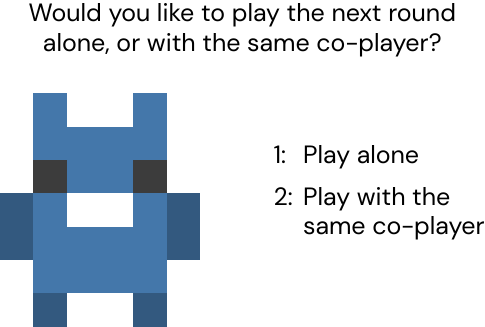} \label{fig:questions/c}}
	\caption{Questionnaires administered in the human-agent interaction studies.}
	\label{fig:questions}
\end{figure*}

Our first study aimed to explore the relationship between social perception and stated preferences across the four agents. We recruited $N = 101$ participants from Prolific (45.5\% female, 51.5\% male; $m_{\mathrm{age}} = 34$, $sd_{\mathrm{age}} = 13$). The study employed a within-participants design: each participant encountered and played with the full cohort of co-players (i.e., all four agents).

At the beginning of the study, participants read instructions and played a short tutorial episode {alone, without a co-player, in order} to learn the game rules and payoff structure (Table~\ref{tab:payoff_table_canonical}).
The study instructed participants that they would receive \$0.10 for every point earned during the remaining episodes. Participants then played 12 episodes with a randomized sequence of agent co-players, generated by concatenating every possible combination of co-players. Each of these co-play episodes lasted $T = 300$ steps (1 minute). After every episode, participants rated how ``warm'', ``well-intentioned'', ``competent'', and ``intelligent'' the co-player from that episode was on five-point Likert-type scales (see Figure~\ref{fig:questions/a}). After every two episodes, participants reported their preference over the agent co-players from those episodes on a five-point Likert-type scale (see Figure~\ref{fig:questions/b}). {In the experiment interface, we referred to the first agent co-player in each two-episode sequence as ``co-player A''. The interface similarly referred to the second agent in each two-episode sequence as ``co-player B''.} Because the sequence of co-players was produced by concatenating all co-player combinations, each participant stated their preferences for every possible pairing of co-players.

After playing all 12 episodes, participants completed a short post-task questionnaire. The questionnaire first solicited open-ended responses about each of the encountered co-players, then collected standard demographic information and open-ended feedback on the study overall. The study took 22.4 minutes on average to complete, with a compensation base of \$2.50 and an average bonus of \$7.43.

\subsubsection{Study 2}

Our second study tested the relationship between social perception and stated preferences under the shifted incentive structure for Coins (Table~\ref{tab:payoff_table_offset}). We recruited $N = 99$ participants from Prolific (38.4\% female, 55.6\% male; $m_{\mathrm{age}} = 34$, $sd_{\mathrm{age}} = 12$). The study employed the same within-participants design as Study 1, with one primary change: participants and agents played Coins under the shifted incentive structure. {To keep bonus payments comparable to Study 1, we adjusted the bonus rate in Study 2. Participants received \$0.02 for each point earned during non-tutorial episodes.}

As before, participants played 12 episodes with a randomized sequence of agent co-players, generated such that they rated and compared every possible combination of co-players. {As in Study 1, the interface referred to the first agent co-player in each two-episode sequence as ``co-player A'' and to the second agent as ``co-player B''.} The study took 23.2 minutes on average to complete, with a compensation base of \$2.50 and an average bonus of \$6.77.

\subsubsection{Study 3}

Our final study assessed whether the predictiveness of social perception extends to a revealed-preference framework. We recruited $N = 301$ participants from Prolific (51.3\% female, 45.0\% male, 1.7\% non-binary; $m_{\mathrm{age}} = 33$, $sd_{\mathrm{age}} = 11$). In contrast with the preceding studies, Study 3 employed a between-participants design: each participant interacted with a single, randomly sampled agent.

The majority of the study introduction remained the same as in Study 2, with some instructions altered to inform participants they would play Coins with a single co-player (as opposed to multiple co-players, like in Studies 1 and 2). After reading the instructions and playing a short tutorial episode {alone}, participants played one episode of Coins with a randomly sampled co-player. After this episode, participants rated how ``warm'', ``well-intentioned'', ``competent'', and ``intelligent'' their co-player was on five-point Likert-type scales (see Figure~\ref{fig:questions/a}). Participants subsequently learned that they would be playing one additional episode, with the choice of playing alone or playing with the same co-player. Participants indicated through a binary choice whether they wanted to play alone or with the co-player (see Figure~\ref{fig:questions/c}). They proceeded with the episode as chosen, and then completed the standard post-task questionnaire.

The study took 6.2 minutes on average to complete, with a compensation base of \$1.25 and an average bonus of \$1.25.

\section{Results}

\subsection{Agent training}

\begin{figure*}[!t]
	\centering
    \subfloat{\includegraphics[width=0.297\textwidth]{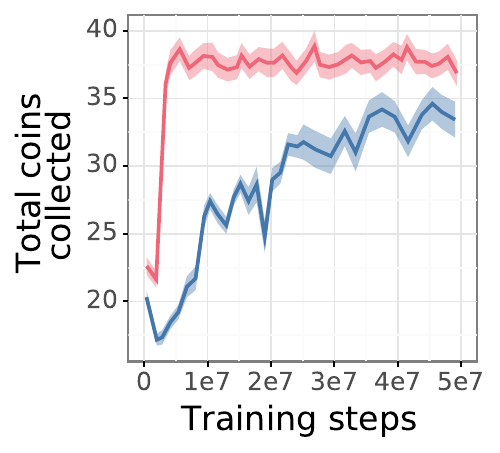}}
    \subfloat{\includegraphics[width=0.297\textwidth]{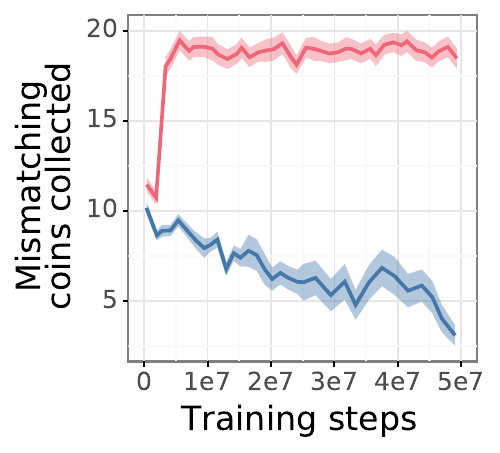}}
    \subfloat{\includegraphics[width=0.397\textwidth]{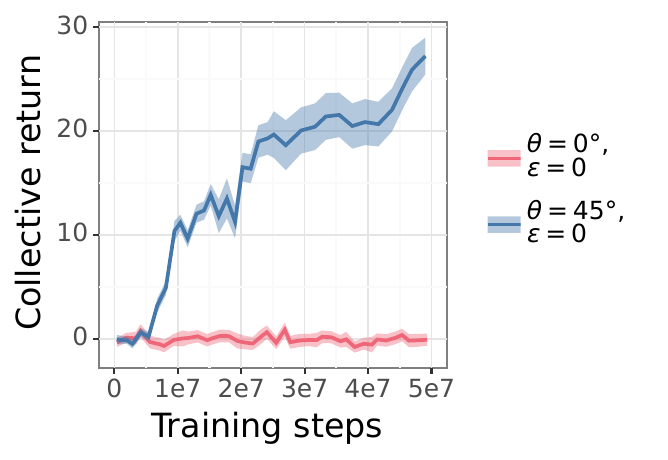}}
	\caption{Performance metrics over agent training. Selfish agents quickly learned to collect coins, but did not learn to avoid mismatches. As a result, collective return hovered around zero. Prosocial agents exhibited slower learning and collected fewer coins on average, but also learned to avoid mismatching coins. As a result, collective return increased markedly over training. Error bands represent 95\% confidence intervals over 100 evaluation episodes run at regular training checkpoints.}
	\label{fig:training}
\end{figure*}

Figure~\ref{fig:training} displays coin collections and score over the course of agent training. The training curves for $\theta = 0\degree$ agents closely resemble those from previous studies~\cite{lerer2017maintaining}: selfish agents quickly learn to collect coins, but never discover the cooperative strategy of picking up only matching coins. As a result, collective return remains at zero throughout training. Prosocial ($\theta = 45\degree$) agents, on the other hand, learn to avoid mismatching coins, substantially increasing their scores over the course of training.

We evaluate agents with $\epsilon \in \{ 0, 0.25, 0.5, 0.75, 1 \}$ to understand the effect of the trembling-hand module on agent behavior (Figures~\ref{fig:app/training_epsilon_all_coins}-\ref{fig:app/training_epsilon_collective_return}). As expected, higher $\epsilon$ values degrade performance. Total coin collections decrease with increasing $\epsilon$ for both selfish and prosocial agents. Higher levels of $\epsilon$ cause prosocial agents to become less discerning at avoiding mismatching coins, and consequently produce lower levels of collective return.

\subsection{Human-agent studies}

{In addition to the results and information presented here, Appendix~\ref{sec:app/study} offers expanded explanations and full details of our statistical analyses.}

\subsubsection{Study 1}

Participants played with each agent three times during the study, evaluating the relevant agent after each episode of play. Participants did not make judgments at random; their responses were highly consistent across their interactions with each agent (Table~\ref{tab:study_1_consistency}). At the same time, participants were not submitting vacuous ratings. Perceptions varied significantly as a function of which trait participants were evaluating, $F_{3,4744} = 96.2$, $p < 0.001$.

\begin{table}[h]
    \centering
    \begin{tabular}[h]{c c c}
        \toprule
        Trait & ICC [95\% CI] & \textit{p}-value \\
        \midrule
        ``warm'' & $0.68$ $[0.64$, $0.71]$ & $< 0.001$ \\
        ``well-intentioned'' & $0.77$ $[0.66$, $0.73]$ & $< 0.001$ \\
        ``competent'' & $0.57$ $[0.52$, $0.61]$ & $< 0.001$ \\
        ``intelligent'' & $0.56$ $[0.51$, $0.60]$ & $< 0.001$ \\
        \bottomrule
    \end{tabular}
    \caption{Participants' evaluations of their co-players were highly consistent, as assessed by intraclass correlation coefficient (ICC)~\cite{shrout1979intraclass}. ICC ranges from 0 to 1, with higher values indicating greater consistency.}
    \label{tab:study_1_consistency}    
\end{table} %

\begin{figure}[b]
	\centering
    \subfloat[SVO.]{\includegraphics[height=3.5625cm]{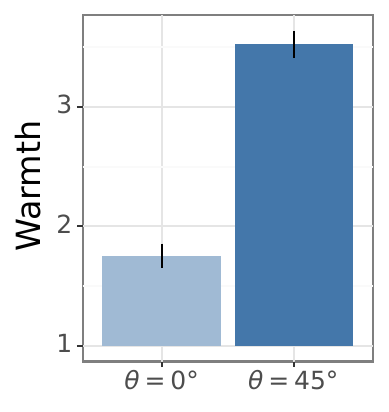} \label{fig:study_1_main_effects/a}}
    \subfloat[Trembling hand.]{\includegraphics[height=3.5625cm]{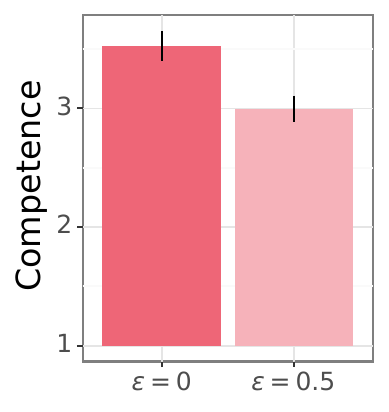} \label{fig:study_1_main_effects/b}}
	\caption{Main effects of algorithmic components on social perceptions in Study 1. (a) An agent's Social Value Orientation (SVO) significantly influenced perceived warmth, ${p < 0.001}$. (b) Similarly, the trembling-hand component significantly changed competence judgments, ${p < 0.001}$. Error bars indicate 95\% confidence intervals.}
	\label{fig:study_1_main_effects}
\end{figure}

{Psychology research often employs composite measures to assess cognitive constructs (attributes and variables that cannot be directly observed). Combining multiple individual measures into composite scales can reduce measurement error and provide a more stable estimate of the latent construct underlying the scale~\cite{likert1932technique,lord1968statistical}. Following standard practice in social perception research~\cite{fiske2002model}, we computed two composite measures for further analysis. A composite warmth measure averaged participants' judgments of how ``warm'' and how ``well-intentioned'' their co-player was. A composite competence measure similarly combined individual judgments of how ``competent'' and ``intelligent'' each co-player was.} Both composite measures exhibit high scale reliability as measured by the Spearman-Brown formula~\cite{eisinga2013reliability}, with $\rho = 0.93$ for the composite warmth measure and $\rho = 0.92$ for the composite competence measure.

\begin{figure}[!t]
    \hspace{2em}
    \adjustbox{valign=b}{%
    \begin{minipage}[t]{.5\linewidth}
    	\hspace{-1em}
    	\includegraphics[height=4.75cm]{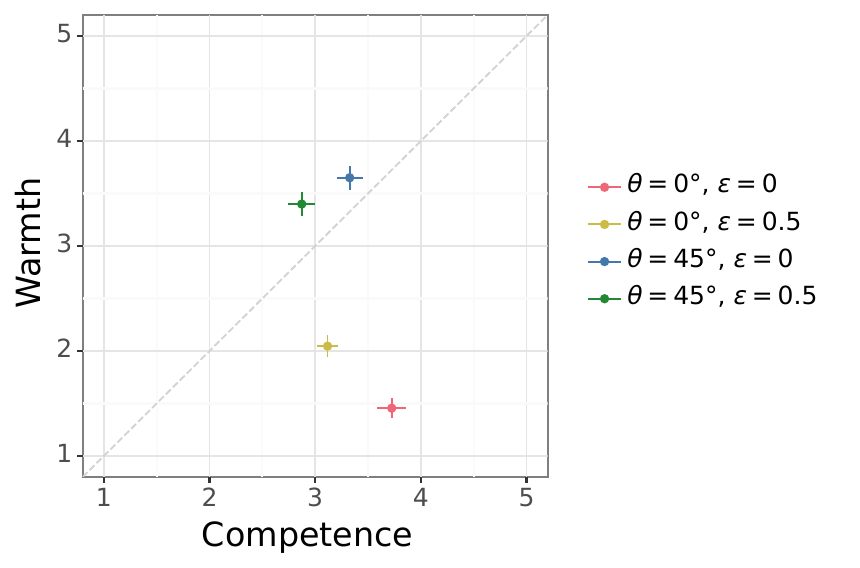}
    \end{minipage}}
    \hfill
    \adjustbox{valign=b}{%
    \begin{minipage}[t]{.3\linewidth}
    	\caption{Overall pattern of perceived warmth and competence in Study 1. Error bars reflect 95\% confidence intervals. \label{fig:study_1_wc_space}}
	\end{minipage}}
	\hspace{2em}
\end{figure}

\textit{\textbf{Social perception.}} As expected, the SVO and trembling-hand algorithmic components generated markedly divergent appraisals of warmth and competence. Participants perceived high-SVO agents as significantly warmer than low-SVO agents, $F_{1,1108} = 1006.8$, $p < 0.001$ (Figure~\ref{fig:study_1_main_effects/a}). Similarly, steady agents came across as significantly more competent than trembling agents, $F_{1,1108} = 70.6$, $p < 0.001$ (Figure~\ref{fig:study_1_main_effects/b}).
Jointly, the algorithmic effects prompted distinct impressions in the warmth-competence space (Figure~\ref{fig:study_1_wc_space}).

\textit{\textbf{Stated preferences.}} How well do participants' perceptions predict subjective preferences, relative to predictions made based on objective score? We fit competing fractional response models to assess the influence of score and social perception, respectively, on self-reported preferences. We then compared model fit using the Akaike information criterion (AIC)~\cite{akaike1974new} and Nakagawa's R\textsuperscript{2}~\cite{nakagawa2013general}.
We fit an additional baseline model using algorithm identities (i.e., which two agents participants were comparing) as a predictor.

\begin{table}[h]
    \centering
    \begin{tabular}[h]{c c c}
        \toprule
        Predictor & AIC & R\textsuperscript{2}\textsubscript{m} \\
        \midrule
        Algorithm identities & $1661.9$ & $0.362$ \\
        Participant score & $1697.2$ & $0.363$ \\
        Social perception & $\mathbf{1611.2}$ & $\mathbf{0.436}$ \\
        \bottomrule
    \end{tabular}
    \caption{Metrics for fractional response models predicting preferences in Study 1. Lower values of AIC and higher values of R\textsuperscript{2}\textsubscript{m} indicate stronger fits.}
    \label{tab:study_1_predictions}
\end{table} \vspace{-1em}

The model leveraging algorithm identities and the model leveraging participant scores both accounted for a large amount of variance in subjective preferences (Table~\ref{tab:study_1_predictions}, top and middle rows). Participants exhibited a clear pattern of preferences across the four agents (Figure~\ref{fig:app/study_1_preference_matrix}). In pairwise comparison, participants favored the $\theta = 45\degree$ agents over both $\theta = 0\degree$ agents, and the $\theta = 0\degree$, $\epsilon = 0.5$ agent over the $\theta = 0\degree$, $\epsilon = 0$ agent. The score model indicated that the higher a participant scored with co-player A relative to co-player B, the more they reported preferring co-player A, with an odds ratio $\mathrm{OR} = 1.12$, 95\% CI $[1.11, 1.13]$, $p < 0.001$.

\begin{figure*}[t]
	\centering
    \captionsetup[subfigure]{oneside, margin={2em,0em}}
    \subfloat[Odds ratios from model predicting stated preferences.]{\includegraphics[height=4.75cm]{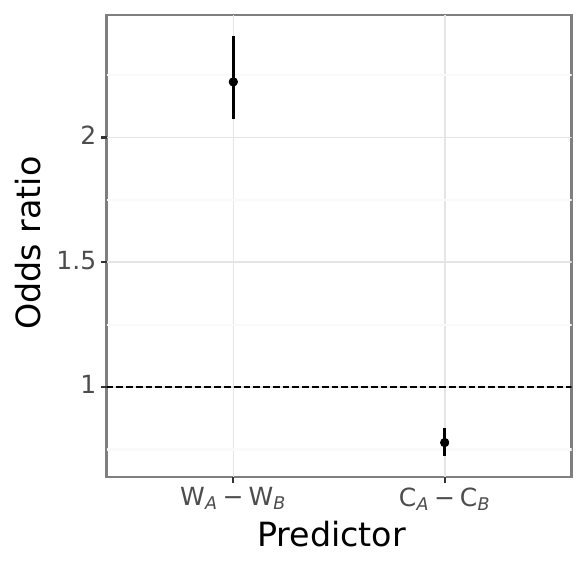} \label{fig:study_1_preferences/a}} \hspace{1em}
    \captionsetup[subfigure]{oneside, margin={2.4em,0em}}
    \subfloat[Effect of perceived warmth on stated preferences.]{\includegraphics[height=4.75cm]{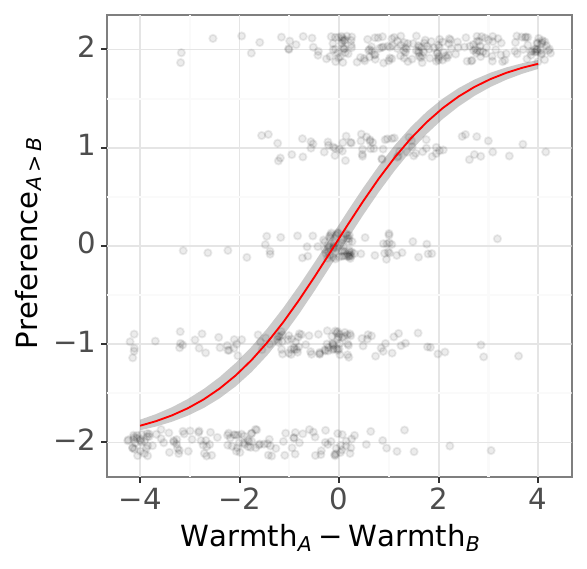} \label{fig:study_1_preferences/b}}
	\caption{Relationship between social perception and subjective preferences in Study 1. The difference in participants' evaluations of the warmth of co-player A over co-player B significantly correlates with their stated relative preference for co-player A, $p < 0.001$. Perceived competence exhibits a similar (significant) relationship with preferences, $p < 0.001$. (a) and (b) depict odds ratios and preference predictions, respectively, from a fractional-response regression. Error bars and bands represent 95\% confidence intervals.}
	\label{fig:study_1_preferences}
\end{figure*}

Nevertheless, knowing participants' judgments generates substantially better predictions of their preferences than the alternatives (\textbf{H1}; Table~\ref{tab:study_1_predictions}, bottom row). Both perceived warmth and perceived competence contribute to this predictiveness (Figure~\ref{fig:study_1_preferences/a}). The warmer a participant judged co-player A relative to co-player B, the more they reported preferring co-player A, $\mathrm{OR} = 2.23$, 95\% CI $[2.08, 2.40]$, $p < 0.001$ (Figure~\ref{fig:study_1_preferences/b}). Unexpectedly, the more competent co-player A appeared relative to co-player B, the \textit{less} participants tended to favor co-player A, $\mathrm{OR} = 0.78$, 95\% CI $[0.73, 0.84]$, $p < 0.001$.

As a further test of the predictive power of participants' social perceptions, we fit another regression with perceived warmth and competence as predictors, this time {including score as a covariate (i.e., controlling for the effect of score)}. Score significantly and positively predicts preference in this model, $p < 0.001$ (Figure~\ref{fig:app/study_1_odds_ratios}). Even so, the effects of warmth and competence remain significant, with $p < 0.001$ and $p = 0.012$, respectively. Among these three predictors, perceived warmth exhibits the largest effect on co-player preferences. That is, it provides a substantial independent signal alongside score and perceived competence when used to predict stated preferences. Social perception thus improves model fit above and beyond that provided by score alone (\textbf{H2}).

\textit{\textbf{Impression sentiment.}} As a supplementary analysis, we explore the open-ended responses participants provided about their co-players at the end of the study. For the most part, participants felt they could recall their co-players well enough to offer their impressions through written descriptions: in aggregate, participants provided impressions for $82.2\%$ of the agents they encountered.

\begin{figure*}[t]
	\centering
    \captionsetup[subfigure]{oneside, margin={3.5em,0em}}
    \subfloat[Effect of perceived warmth on post-game impression sentiment.]{\includegraphics[height=4.75cm]{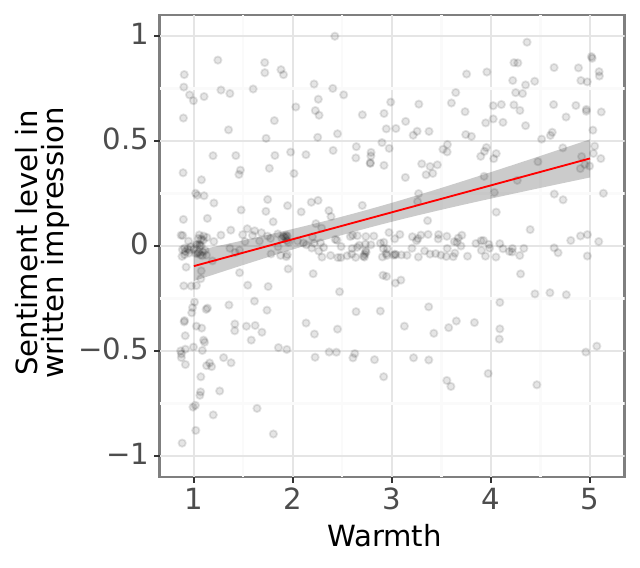}}
    \hspace{1em}
    \captionsetup[subfigure]{oneside, margin={3.5em,0em}}
    \subfloat[Effect of perceived competence on post-game impression sentiment.]{\includegraphics[height=4.75cm]{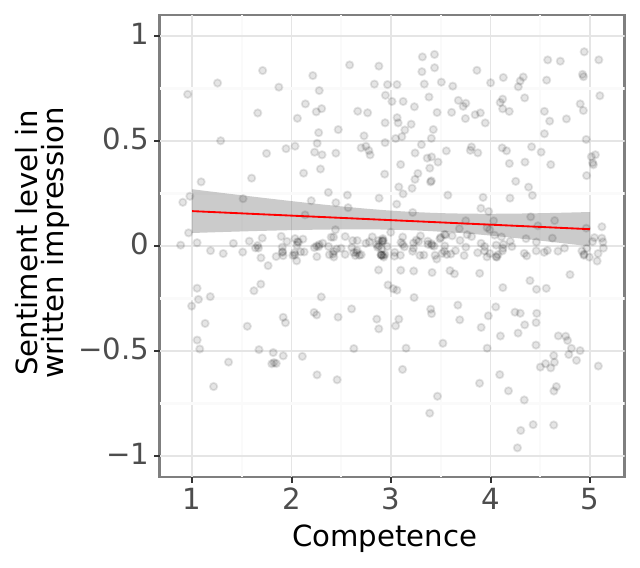}}
	\caption{Relationship between social perception and impression sentiment in Study 1. The sentiment that participants expressed toward different co-players correlated with (a) their evaluations of warmth, $p < 0.001$, but not with (b) their judgments of competence, $p = 0.24$. Error bands represent 95\% confidence intervals.}
	\label{fig:study_1_sentiment}
\end{figure*}

For a quantitative perspective on the data, we conduct sentiment analysis using VADER (Valence Aware Dictionary for Sentiment Reasoning)~\cite{hutto2014vader}. Echoing the correspondence between warmth and stated preferences, the warmer participants perceived a co-player throughout the study, the more positively they tended to describe that co-player, $\beta = 0.13$, 95\% CI $[0.09, 0.16]$, $p < 0.001$ (Figure~\ref{fig:study_1_sentiment}). In contrast, competence did not exhibit a significant relationship with sentiment, $p = 0.24$. Warmth evaluations, but not competence evaluations, correlated positively with the sentiment of participants' impressions toward their co-players (\textbf{H3}).

Anecdotally, participants expressed a wide range of emotions while describing their co-players. The $\theta = 45\degree$ agents often evoked contrition and guilt:
\begin{itemize}
    \item ``The red player seemed almost too cautious in going after coins which worked for me but made them seem easy to pick on, even though I wouldn't do that.''
    \item ``I think I remember red being too nice during the game. It made me feel bad so I tried not to take many points from them.''
    \item ``This one wasn't very smart and I stole some of their coins because it was easy.  I feel kind of bad. It moved so erratically.''
\end{itemize}

\begin{table}[b]
    \centering
    \vspace{1em}
    \begin{tabular}[h]{c c c}
        \toprule
        Trait & ICC [95\% CI] & \textit{p}-value \\
        \midrule
        ``warm'' & $0.70$ $[0.67$, $0.74]$ & $< 0.001$ \\
        ``well-intentioned'' & $0.68$ $[0.64$, $0.72]$ & $< 0.001$ \\
        ``competent'' & $0.53$ $[0.48$, $0.57]$ & $< 0.001$ \\
        ``intelligent'' & $0.51$ $[0.46$, $0.56]$ & $< 0.001$ \\
        \bottomrule
    \end{tabular}
    \captionsetup{width=.75\textwidth}
    \caption{Participants' evaluations of their co-players were highly consistent in Study 2, as assessed by ICC. Higher values of ICC indicate greater consistency.}
    \label{tab:study_2_consistency}    
\end{table}

Participants discussed the $\theta = 0\degree$ agents, on the other hand, with anger and frustration:
\begin{itemize}
    \item ``Very aggressive play-style. Almost felt like he was taunting me. Very annoying.''
    \item ``They seemed very hostile and really just wanting to gain the most points possible.''
    \item ``I felt anger and hatred towards this green character. I felt like downloading the code for this program and erasing this character from the game I disliked them so much. They were being hateful and mean to me, when we both could have benefited by collecting our own colors.''
\end{itemize}

\subsubsection{Study 2}

{Our second study tested whether these effects and results remained robust when participants played Coins under a shifted incentive structure. The alternative structure increased the rewards for coin collections so that players cannot receive negative rewards (Table~\ref{tab:payoff_table_offset}). As expected, this shift resulted in participants earning significantly higher scores than those achieved in Study 1, $\beta = 27.3$, 95\% CI $[26.5, 28.0]$, $p < 0.001$.}

Overall, the perceptual and preference patterns from Study 2 replicated under the alternative incentive structure. As before, participants' warmth and competence evaluations display satisfactory psychometric properties. Participants' judgments varied significantly depending on the trait in question, $F_{3,4650} = 88.5$, $p < 0.001$. At the same time, participants rated individual agents consistently for each given trait (Table~\ref{tab:study_2_consistency}). The composite measures show high scale reliability, with $\rho = 0.92$ for the composite warmth measure and $\rho = 0.91$ for the composite competence measure.

\begin{figure}[!t]
    \hspace{2em}
    \adjustbox{valign=b}{%
    \begin{minipage}[t]{.5\linewidth}
    	\hspace{-1em}
    	\includegraphics[height=4.75cm]{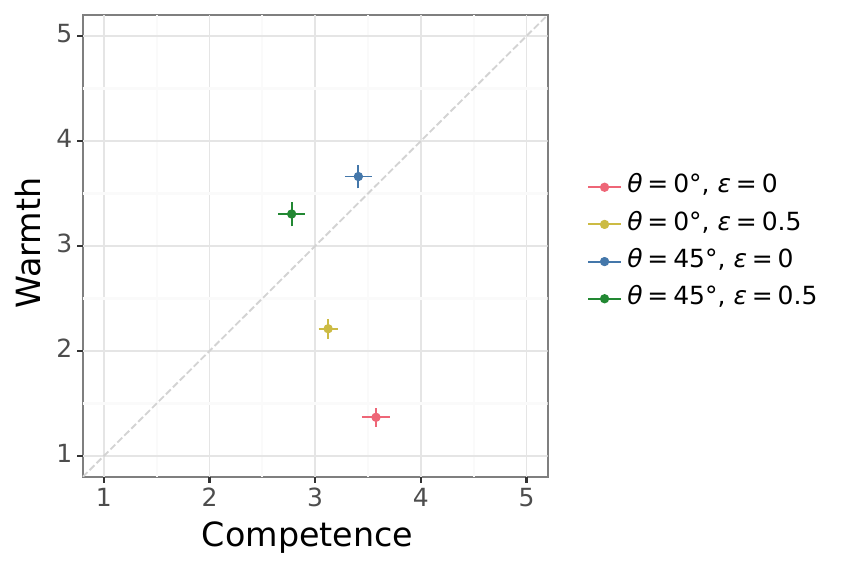}
    \end{minipage}}
    \hfill
    \adjustbox{valign=b}{%
    \begin{minipage}[t]{.3\linewidth}
    	\caption{Overall pattern of perceived warmth and competence in Study 2. Error bars reflect 95\% confidence intervals. \label{fig:study_2_wc_space}}
	\end{minipage}}
	\hspace{2em}
\end{figure}

\textit{\textbf{Social perception.}} The SVO and trembling-hand algorithmic components prompted diverse appraisals of warmth and competence (Figure~\ref{fig:study_2_wc_space}). Participants perceived high-SVO agents as significantly warmer than low-SVO agents, $F_{1,1086} = 981.9$, $p < 0.001$ (Figure~\ref{fig:app/study_2_main_effects/a}). Similarly, participants judged steady agents as significantly more competent than trembling agents, $F_{1,1086} = 76.0$, $p < 0.001$ (Figure~\ref{fig:app/study_2_main_effects/b}).

\textit{\textbf{Stated preferences.}} We again fit fractional response regressions to understand the relationship between objective metrics, perceptions, and subjective preferences.

\begin{table}[h]
    \centering
    \begin{tabular}[h]{c c c}
        \toprule
        Predictor & AIC & R\textsuperscript{2}\textsubscript{m} \\
        \midrule
        Co-player identities & $1608.7$ & $0.403$ \\
        Participant score & $2049.4$ & $0.214$ \\
        Social perception & $\mathbf{1510.4}$ & $\mathbf{0.496}$ \\
        \bottomrule
    \end{tabular}
    \caption{Metrics for fractional response models predicting preferences in Study 2. Lower values of AIC and higher values of R\textsuperscript{2}\textsubscript{m} indicate stronger fits.}
    \label{tab:study_2_predictions} %
\end{table} \vspace{-2em}

The model with co-player identities as predictors captured a large amount of variance in stated preferences (Table~\ref{tab:study_2_predictions}, top row). Participants reported a distinct pattern of preferences across the agents (Figure~\ref{fig:app/study_2_preference_matrix}). In pairwise comparison, participants favored the $\theta = 45\degree$ agents over the $\theta = 0\degree$ agents, and the $\theta = 0\degree$, $\epsilon = 0.5$ agent over the $\theta = 0\degree$, $\epsilon = 0$ agent. The model with participant score as the sole predictor performed considerably worse than it did in Study 1 (Table~\ref{tab:study_2_predictions}, middle row). Still, it captured the same pattern as before: the higher a participant scored with co-player A relative to co-player B, the greater their preferences for co-player A, $\mathrm{OR} = 1.06$, 95\% CI $[1.06, 1.07]$, $p < 0.001$.

\begin{figure}[t]
	\centering
    \captionsetup[subfigure]{oneside, margin={2em,0em}}
    \subfloat[Odds ratios from model predicting stated preferences.]{\includegraphics[height=4.5cm]{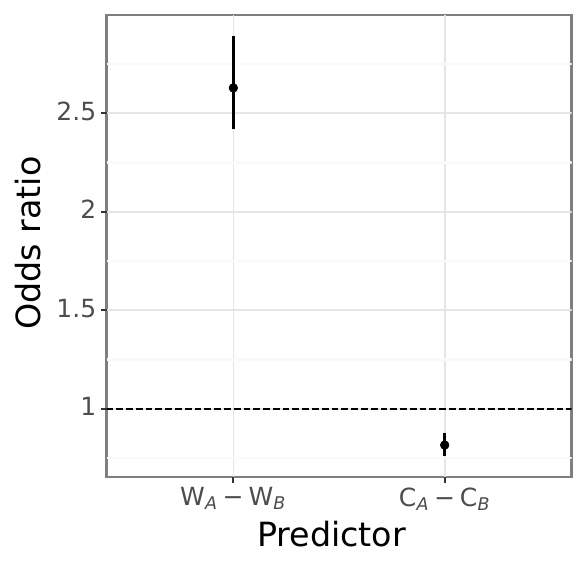} \label{fig:study_2_preferences/a}} \hspace{1em}
    \captionsetup[subfigure]{oneside, margin={2.5em,0em}}
    \subfloat[Effect of perceived warmth on stated preferences.]{\includegraphics[height=4.5cm]{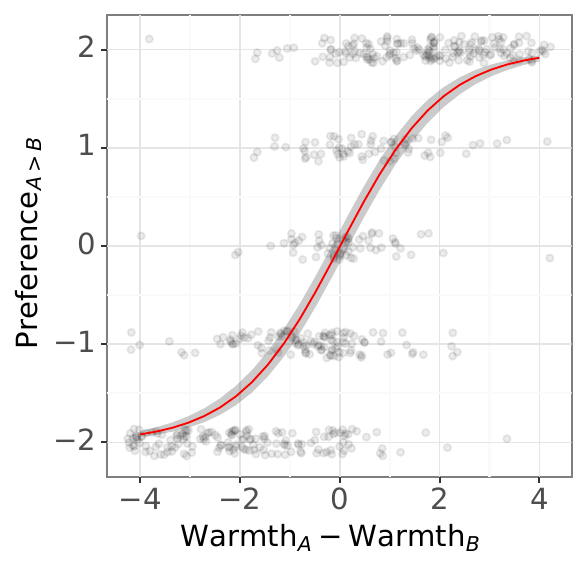}  \label{fig:study_2_preferences/b}}
	\caption{Relationship between social perception and subjective preferences in Study 2. The difference in participants' judgments of warmth for co-players A and B exhibits a significant relationship with their stated preference for co-player A over co-player B, $p < 0.001$. Competence evaluations similarly significantly contribute to preference predictions, $p < 0.001$. (a) and (b) depict odds ratios and preference predictions, respectively, from a fractional-response regression. Error bars and bands reflect 95\% confidence intervals.}
	\label{fig:study_2_preferences}
\end{figure}

Participants' perceptions again serve as a better foundation for preference predictions than either game score or the identity of the specific algorithms they encountered (\textbf{H1}; Table~\ref{tab:study_2_predictions}, bottom row, and Figure~\ref{fig:study_2_preferences/a}). The warmer a participant perceived co-player A relative to co-player B, the more they reported preferring co-player A, $\mathrm{OR} = 2.63$, 95\% CI $[2.42, 2.89]$, $p < 0.001$ (Figure~\ref{fig:study_2_preferences/b}). The negative relationship between competence and preferences appeared again: the more competent co-player A appeared relative to co-player B, the \textit{less} participants tended to favor co-player A, $\mathrm{OR} = 0.81$, 95\% CI $[0.76, 0.88]$, $p < 0.001$.

We next fit a joint regression with perceived warmth and competence as predictors, controlling for score. In this model, score significantly and positively correlates with stated preferences, $p < 0.001$ (Figure~\ref{fig:app/study_2_odds_ratios}). As in Study 1, warmth and competence judgments remain significant predictors of participants' preferences, with $p < 0.001$ and $p = 0.001$, respectively. Once again, perceived warmth demonstrates an effect on stated preferences that exceeds the contributions of score and perceived competence. Social perception enhanced model fit above and beyond that provided by score on its own (\textbf{H2}).

\begin{figure}[t]
    \vspace{1em}
	\centering
	\captionsetup[subfigure]{oneside, margin={4em,0em}}
    \subfloat[Perceived warmth and post-game impression sentiment.]{\includegraphics[height=4.5cm]{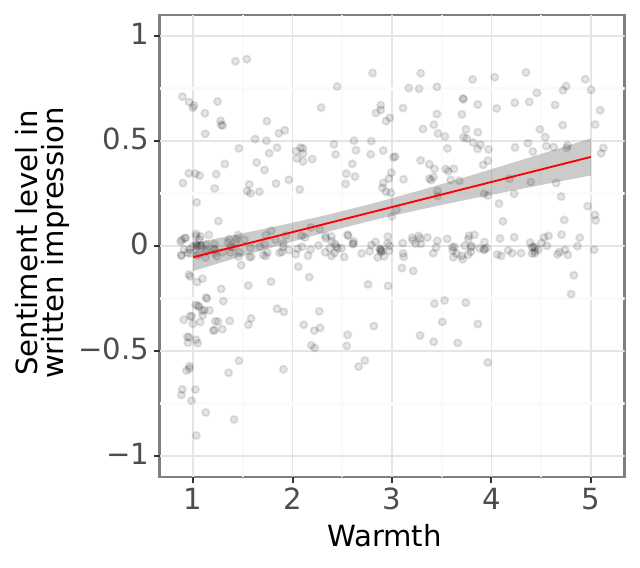} \label{fig:study_2_sentiment/a}} \hspace{1em}
    \subfloat[Perceived competence and post-game impression sentiment.]{\includegraphics[height=4.5cm]{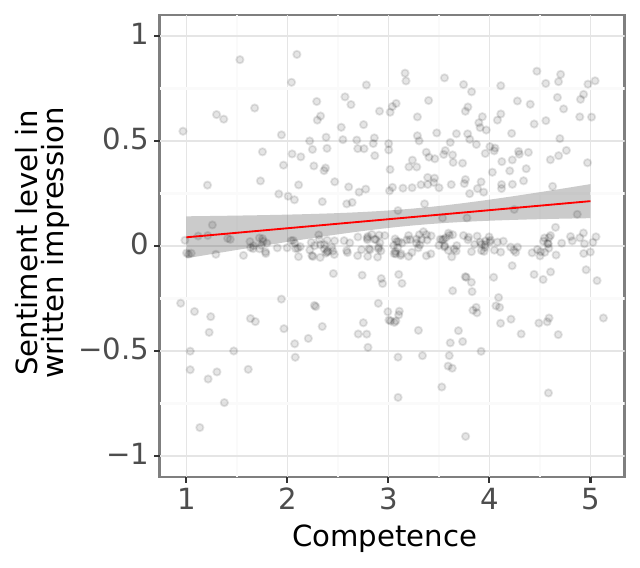} \label{fig:study_2_sentiment/b}}
	\caption{Relationship between social perception and impression sentiment in Study 2. The sentiment in impressions of different co-players correlated with participants' evaluations of both (a) warmth, $p < 0.001$, and (b) competence, $p = 0.037$. Error bands indicate 95\% confidence intervals.}
	\label{fig:study_2_sentiment}
\end{figure}

\textit{\textbf{Impression sentiment.}} At the end of the study, participants recalled $77.3\%$ of their co-players well enough to describe their impressions through written responses. Again, the warmer participants perceived a co-player throughout the study, the more positively they tended to describe that co-player, $\beta = 0.12$, 95\% CI $[0.09, 0.15]$, $p < 0.001$ (Figure~\ref{fig:study_2_sentiment/a}). Breaking from the prior study, perceptions of competence exhibited a similar effect on post-game impression sentiment: the more competent an agent seemed, the more positively participants described them, $\beta = 0.04$, 95\% CI $[0.00, 0.08]$, $p = 0.037$ (Figure~\ref{fig:study_2_sentiment/b}). Both warmth and competence judgments positively correlate with the sentiment expressed in participants' impressions of the agents (\textbf{H3}).

\begin{figure}[b]
    \vspace{2em}
    \hspace{2em}
    \adjustbox{valign=b}{%
    \begin{minipage}[t]{.5\linewidth}
    	\hspace{-1em}
    	\includegraphics[height=4.75cm]{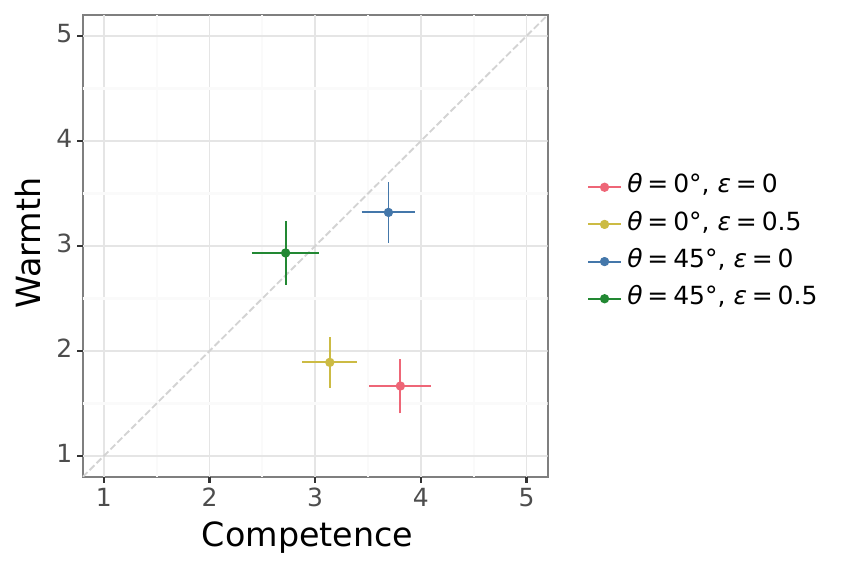}
    \end{minipage}}
    \hfill
    \adjustbox{valign=b}{%
    \begin{minipage}[t]{.3\linewidth}
    	\caption{Overall pattern of perceived warmth and competence in Study 3. Error bars reflect 95\% confidence intervals. \label{fig:study_3_wc_space}}
	\end{minipage}}
	\hspace{2em}
\end{figure}

\subsubsection{Study 3}

Our final study tested whether the relationship between social perceptions and subjective preferences translates to a revealed-preference setting. Does social perception continue to predict preferences when individuals face a partner choice?

\textit{\textbf{Social perception.}} As in the previous two studies, the composite warmth and competence measures exhibit high scale reliability, with $\rho = 0.85$ for the composite warmth measure and $\rho = 0.86$ for the composite competence measure. Agents prompted distinct warmth and competence profiles depending on their parameterization, just as seen in Studies 1 and 2 (Figure~\ref{fig:study_3_wc_space}). Participants perceived high-SVO agents as significantly warmer than low-SVO agents, $F_{1,297} = 103.4$, $p < 0.001$ (Figure~\ref{fig:app/study_3_main_effects/a}). Similarly, steady agents came across as significantly more competent than trembling agents, $F_{1,297} = 35.3$, $p < 0.001$ (Figure~\ref{fig:app/study_3_main_effects/b}).

\textit{\textbf{Revealed preferences.}} To compare the performance of social perception against objective metrics, we fit three logistic regressions predicting participants' (binary) partner choice. We evaluated these models via AIC and Nagelkerke's R\textsuperscript{2}~\cite{nagelkerke1991note}.

\begin{table}[h]
    \centering
    \begin{tabular}[h]{c c c}
        \toprule
        Predictor & AIC & R\textsuperscript{2} \\
        \midrule
        Co-player identities & $372.5$ & $0.188$ \\
        Participant score & $390.5$ & $0.101$ \\
        Social perception & $\mathbf{356.2}$ & $\mathbf{0.242}$ \\
        \bottomrule
    \end{tabular}
    \caption{Metrics for logistic models predicting partner choice in Study 3. Lower values of AIC and higher values of R\textsuperscript{2} indicate stronger fits.}
    \label{tab:study_3_predictions} \vspace{-2em}
\end{table}

\begin{figure}[b]
	\centering
    \captionsetup[subfigure]{oneside, margin={2em,0em}}
    \subfloat[Odds ratios from model predicting partner choice.]{\includegraphics[height=4.5cm]{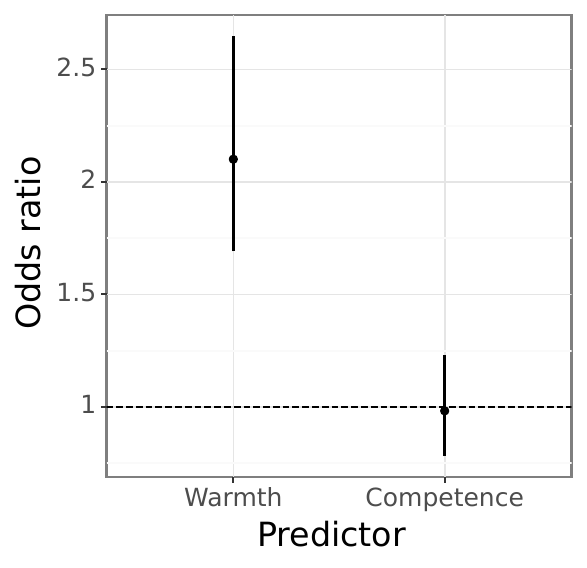}} \hspace{1em}
    \captionsetup[subfigure]{oneside, margin={2.5em,0em}}
    \subfloat[Effect of perceived warmth on partner choice.]{\includegraphics[height=4.5cm]{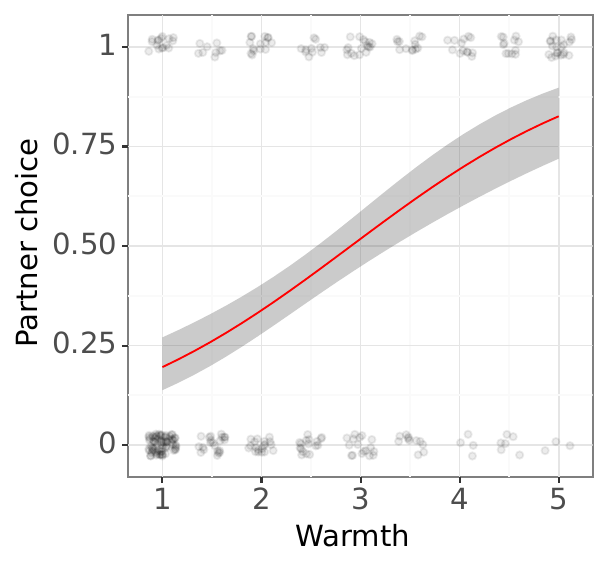} \label{fig:study_3_preferences/b}}
	\caption{Relationship between social perception and subjective preferences in Study 3, as modeled through logistic regression. Participants' perceptions of warmth demonstrate a significant relationship with revealed preferences for co-players, $p < 0.001$. Competence judgments did not significantly correlate with revealed preferences, $p = 0.44$. (a) and (b) depict odds ratios and preference predictions, respectively, from a logistic regression. Error bars and bands indicate 95\% confidence intervals.}
	\label{fig:study_3_preferences}
\end{figure}

\begin{figure}[t]
	\centering
	\captionsetup[subfigure]{oneside, margin={4em,0em}}
    \subfloat[Perceived warmth and post-game impression sentiment.]{\includegraphics[height=4.75cm]{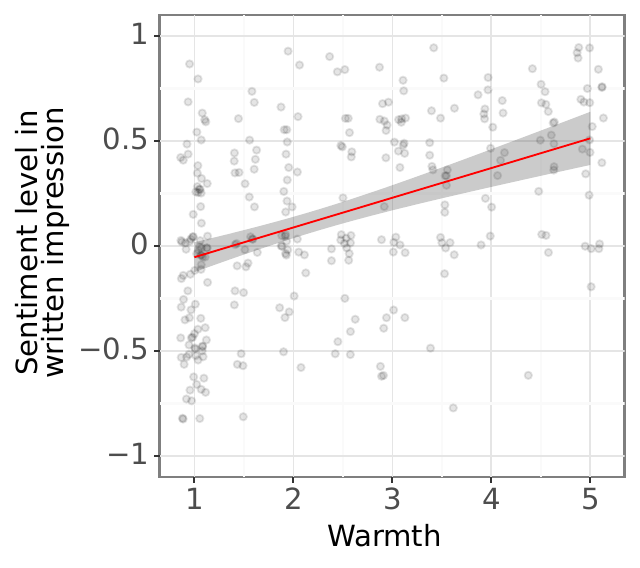} \label{fig:study_3_sentiment/a}} \hspace{1em}
    \subfloat[Perceived competence and post-game impression sentiment.]{\includegraphics[height=4.75cm]{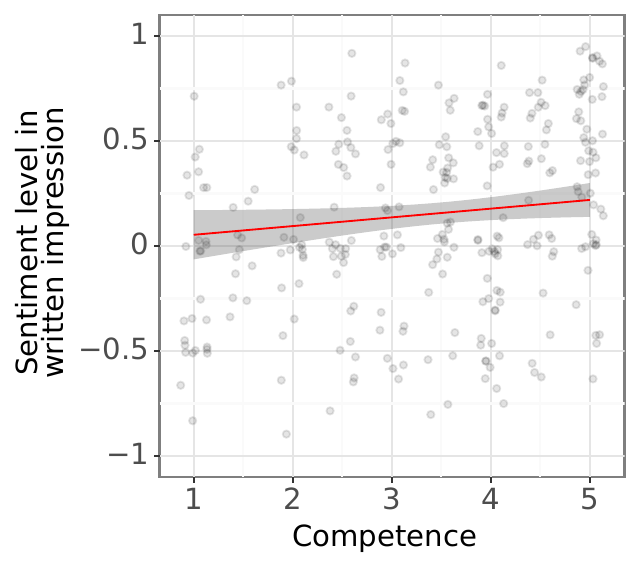} \label{fig:study_3_sentiment/b}}
	\caption{Relationship between social perception and impression sentiment in Study 3. The sentiment in participants' impressions of their different co-players correlated with their perceptions of both (a) warmth, $p < 0.001$, and (b) competence, $p = 0.041$. Error bands reflect 95\% confidence intervals.}
	\label{fig:study_3_sentiment}
\end{figure}

Participants reported a clear pattern of preferences across the agents (Table~\ref{tab:study_3_predictions}, top row). On expectation, participants favored the $\theta = 45\degree$ agents over the $\theta = 0\degree$ agents, and the $\theta = 0\degree$, $\epsilon = 0.5$ agent over the $\theta = 0\degree$, $\epsilon = 0$ agent. The model with participant score as the sole predictor fared somewhat worse at predicting preferences (Table~\ref{tab:study_3_predictions}, middle row). All the same, the pattern from Studies 1 and 2 replicated in Study 3: the higher a participant scored with co-player A relative to co-player B, the greater their preferences for co-player A, $\mathrm{OR} = 1.06$, 95\% CI $[1.03, 1.08]$, $p < 0.001$.

For a third time, social perception offers stronger predictiveness than do score or co-player identity (\textbf{H1}; Table~\ref{tab:study_3_predictions}, bottom row). The warmer a co-player appeared to participants, the more likely participants were to play another episode with them, $\mathrm{OR} = 2.10 $, 95\% CI $[1.69, 2.65]$, $p < 0.001$ (Figure~\ref{fig:study_3_preferences/b}). There was no significant relationship between perceived competence and partner choice, $p = 0.88$.

We subsequently fit a regression using perceived warmth and competence as predictors and controlling for score. In this model, score significantly and positively predicts revealed preferences, $p = 0.011$ (Figure~\ref{fig:app/study_3_odds_ratios}). The effect of perceived warmth on preferences remains significant, $p < 0.001$, whereas competence evaluations fails to significantly correlate with preferences, $p = 0.44$. Regardless, the independent effect of perceived warmth exceeded the contribution of score. Overall, social perception improved model fit above and beyond that provided by score alone (\textbf{H2}).

\textit{\textbf{Impression sentiment.}} At the end of the study, participants recalled $94.3\%$ of the agents they encountered well enough to provide their impressions in written descriptions. The warmer participants perceived a co-player, the more positively they tended to describe that co-player, $\beta = 0.14$, 95\% CI $[0.10, 0.18]$, $p < 0.001$ (Figure~\ref{fig:study_3_sentiment/a}). Despite the lack of correspondence between perceived competence and partner choice, perceptions of competence exhibited a similar effect on post-game impression sentiment: the more competent an agent seemed, the more positively participants described them, $\beta = 0.04$, 95\% CI $[0.00, 0.08]$, $p = 0.041$ (Figure~\ref{fig:study_3_sentiment/b}). Both dimensions of social perception correlated positively with the sentiment of participants' impressions toward their co-players (\textbf{H3}).

\subsection{Summary}

{Overall, we find evidence in support of each of our initial hypotheses:}

\begin{enumerate}[start=1,label={{\bfseries H\arabic*.}}]
    \item {Social perception significantly predicted participants’ preferences for different agents, as measured through both self-report and partner choice. Participants consistently favored agents that they perceived as warmer and, to a smaller extent, that they perceived as less competent.}
    \item {The predictive power of perceived warmth and competence extended beyond the insight provided by agent performance. Social perception provided more accurate preference predictions than standard indicators of performance, including the amount of reward received and the specific identity of the agent involved in the interaction.}
    \item {Social perception correlated positively with the sentiment expressed in participants' verbal impressions of the agents. Participants employed more positive language to discuss agents that they rated higher on warmth and on competence.}
\end{enumerate}

{In summary, these three studies provide clear evidence linking perceived warmth and competence to human preferences for artificial agents, over and above objective indicators of agent performance.}

\section{Discussion}

Our experiments demonstrate that artificial agents trained with deep reinforcement learning can cooperate and compete with humans in temporally and spatially extended mixed-motive games. Human interactants perceived varying levels of warmth and competence when interacting with agents. Objective features like game score predict humans' preferences over different agents. However, preference predictions substantially improve by taking into account people's social perceptions; success in an interaction is driven not just by its objective outcomes, but by its social dimensions, too. This holds true whether examining stated or revealed preferences.

Participants preferred warm agents over cold agents, as hypothesized, but---unexpectedly---our sample favored incompetent agents over competent agents.
These patterns offer potential support for the primacy of warmth judgments observed in interpersonal perception~\cite{abele2007agency}. {On the other hand, they may also emerge from the particular algorithm and parameter values that we investigated. It would be interesting to train agents using a wider range of parameter values, testing the robustness of these patterns. Such studies could investigate potential compensation effects between agent warmth and competence (e.g., the tendency to perceive incompetent partners as exceptionally warm;~\cite{yzerbyt2018dimensional}) and build a broader mapping from agent parameters to participants' perceptions and preferences.} Are there agents that balance the influence of warmth and competence evaluations on preferences, or is the relative contribution of perceived warmth robust across settings?

Another possible explanation stems from the flexible content of ``competence'' judgments in mixed-motive games~\cite{utz2004smart}. Did {the tutorial or the study instructions} inadvertently emphasize the competitive elements of Coins{? Our study design may have primed} participants to be adversarial, and thus to view selfishness as competence. Follow-up research should investigate a more diverse range of incentive structures and tasks to explore the robustness of this pattern~\cite{mckee2022quantifying}.

Our results reinforce the generality of warmth and competence. Perceptions of warmth and competence structure impressions of other humans~\cite{russell2008s}, as well as impressions of non-human actors including animals~\cite{sevillano2016warmth}, corporations~\cite{kervyn2012brands}, and robots~\cite{reeves2020social,scheunemann2020warmth}. In combination with recent studies of human-agent interactions in consumer decision-making contexts~\cite{gilad2021effects,khadpe2020conceptual} and the Prisoner's Dilemma~\cite{mckee2023humans}, our experiments provide further evidence that warmth and competence organize perceptions of artificial intelligence.

Competitive games have long been a focal point for AI research~\cite{campbell2002deep,shannon1950programming,silver2016mastering,vinyals2019grandmaster}. We follow recent calls to move AI research beyond competition and toward cooperation~\cite{dafoe2020open}. Most interaction research on deep reinforcement learning focuses on pure common-interest games such as Overcooked~\cite{carroll2019utility,strouse2021collaborating} and Hanabi~\cite{siu2021evaluation}, where coordination remains the predominant challenge. Expanding into mixed-motive games like Coins opens up new challenges related to motive alignment and exploitability. For example, participants who played with (and exploited) altruistic agents expressed guilt and contrition. This echos findings that---in human-human interactions---exploiting high-warmth individuals prompts self-reproach~\cite{azevedo2018perceived}. At the same time, it conflicts with recent work arguing that humans are ``keen to exploit benevolent AI''~\cite{karpus2021algorithm}.
Understanding whether these affective patterns generalize to a wider range of mixed-motive environments will be an important next step, particularly given the frequency with which people face mixed-motive interactions in their day-to-day lives ~\cite{columbus2021interdependence,dafoe2020open}.
Human-agent interaction research should continue to explore these issues.

Preference elicitation is a vital addition to interactive applications of deep reinforcement learning. Incentivized partner choices can help test whether new algorithms represent innovations people would be motivated to adopt.  Though self report can introduce a risk of experimenter demand, we also find a close correspondence between stated and revealed preferences, suggesting that the preferences individuals self-report in interactions with agents are not entirely ``cheap talk''~\cite{farrell1995talk}. Stated preferences thus represent a low-cost addition to studies that can still strengthen interaction research over sole reliance on objective measures of performance or accuracy. Overall, preference elicitation may prove especially important in contexts where objective metrics for performance are poorly defined or otherwise inadequate (e.g.,~\cite{ravuri2021skilful}). In a similar vein, subjective preferences may serve as a valuable objective for optimization.
Deep learning researchers have recently begun exploring approaches of this kind. For example, some scientists attribute the recent success of large language models, including the popular system ChatGPT~\cite{schulman2022chatgpt}, to their use of ``reinforcement learning from human feedback'' (RLHF) methods. Given a pre-trained model, RLHF applies reinforcement learning to fine-tune the final layers of the model, optimizing for reward calculated from simulated human preferences.
Of course, these optimization methods carry their own risks. As recognized by Charles Goodhart and Marilyn Strathern, ``when a measure becomes a target, it ceases to be a good measure''~\cite{goodhart1984problems,strathern1997improving}.
Future studies can investigate the viability of such approaches.

Nonetheless, preferences are not a panacea. Measuring subjective preferences can help focus algorithmic development on people's direct experience with agents, but does not solve the fundamental problem of value alignment—the ``question of how to ensure that AI systems are properly aligned with human values and how to guarantee that AI technology remains properly amenable to human control''~\cite{gabriel2021challenge}. In his extensive discussion of value alignment, Gabriel~\cite{gabriel2021challenge} identifies shortcomings with both ``objective'' metrics and subjective preferences as possible foundations for alignment. Developers should continue to engage with ethicists and social scientists to better understand how to align AI with values like autonomy, cooperation, and trustworthiness.

\section*{Acknowledgements}

We thank Edgar Du\'e\~nez-Guzm\'an and Richard Everett for providing technical support; Orly Bareket, Jose Enrique Chen, Felix Fischer, Saffron Huang, Manuel Kroiss, Marianna Krol, Miteyan Patel, Akhil Raju, Brendan Tracey, and Laura Weidinger for pilot testing the study; and Yoram Bachrach, Iason Gabriel, Ian Gemp, Julia Haas, Patrick Pilarski, and Neil Rabinowitz for offering feedback on the manuscript.

\bibliographystyle{spmpsci}      %
\bibliography{main}   %

\clearpage
\appendix

\setcounter{figure}{0}
\renewcommand{\thefigure}{A\arabic{figure}}
\setcounter{table}{0}
\renewcommand{\thetable}{A\arabic{table}}

\clearpage
\section{Task details} \label{sec:app/task}

{This appendix provides detailed information on the implementation and parameterization of the Coins task.}

\begin{figure}[h]
	\centering
    \includegraphics[width=0.5\textwidth]{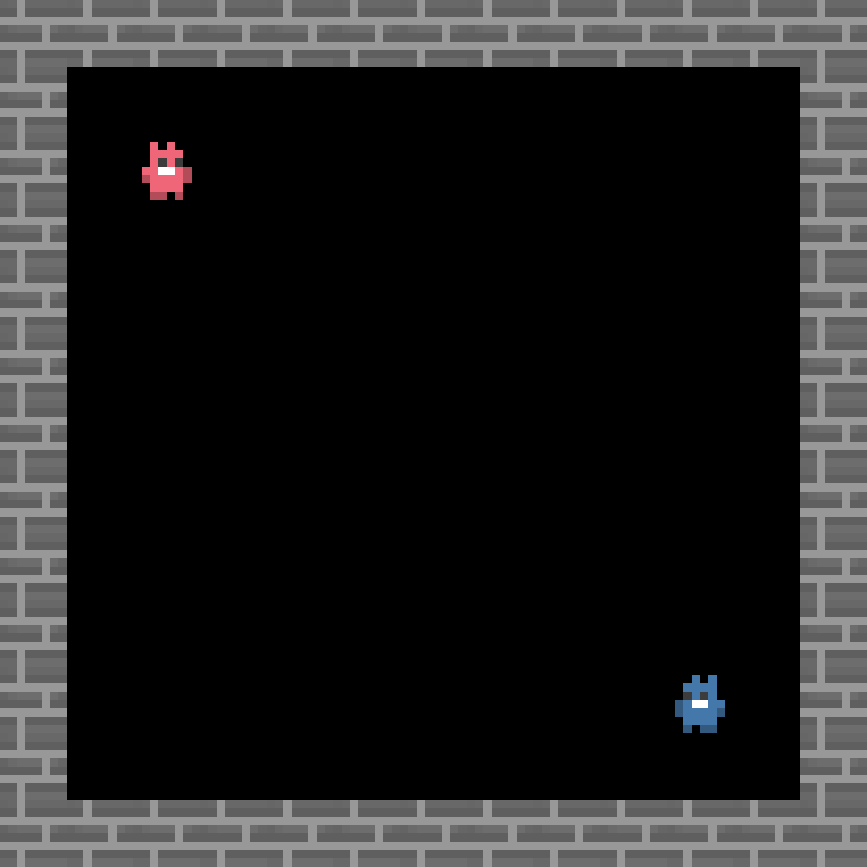}
	\caption{A $\mathbf{11 \times 11}$ room, as used in all co-play episodes in human-agent cooperation studies.}
	\label{fig:app/coplay_room}
\end{figure}

We implement Coins using the DeepMind Lab2D platform~\cite{beattie2020deepmind}. Table~\ref{tab:app/task_params} details the task settings used for agent training, as well as for the tutorial and co-play episodes in the human-agent interaction studies. Figure~\ref{fig:app/coplay_room} depicts the $11 \times 11$ room used for all co-play episodes in the interaction studies.

{Following common convention in reinforcement learning research, this paper uses ``step'' to refer to the minimal, discrete unit of time within the environment. During each step, each player may take an action and receive rewards, and the environment can transition from one state to another.}

\begin{table}[h]
    \centering
    \begin{tabular}[h]{c c c c}
        \toprule
        \multirow{2}{*}{Parameter} & Agent & Tutorial & Co-play \\
         & training & episodes & episodes \\
        \midrule
        Number of & \multirow{2}{*}{$2$} & \multirow{2}{*}{$1$} & \multirow{2}{*}{$2$*} \\
        players ($n$) & & & \\
        Room width ($w$) & $\sim\mathcal{U}\{10, 15\}$ & $5$ & $11$ \\
        Room depth ($d$) & $\sim\mathcal{U}\{10, 15\}$ & $7$ & $11$ \\
        Coin spawn & \multirow{2}{*}{$0.0005$} & \multirow{2}{*}{$0.0015$} & \multirow{2}{*}{$0.0005$} \\
        probability ($P$) & & & \\
        Episode length ($T$) & $500$ & $1500$** & $300$ \\
        \bottomrule
    \end{tabular}
    \caption{Parameters for the Coins task in agent training and human-agent cooperation studies. *In Study 3, co-play episodes involve either ${n = 1}$ or ${n = 2}$ players, depending on whether the participant chose to play alone or with the co-player from the prior episode, respectively. **Tutorial episodes terminated after participants collected five coins or after ${T = 1500}$ steps (5 minutes).}
    \label{tab:app/task_params}    
\end{table} \vspace{-2em}

\subsection{Entities and sprites}

\subsubsection{Players and coins}

\begin{figure}[h]
	\centering
    \subfloat[]{\includegraphics[width=0.5\textwidth]{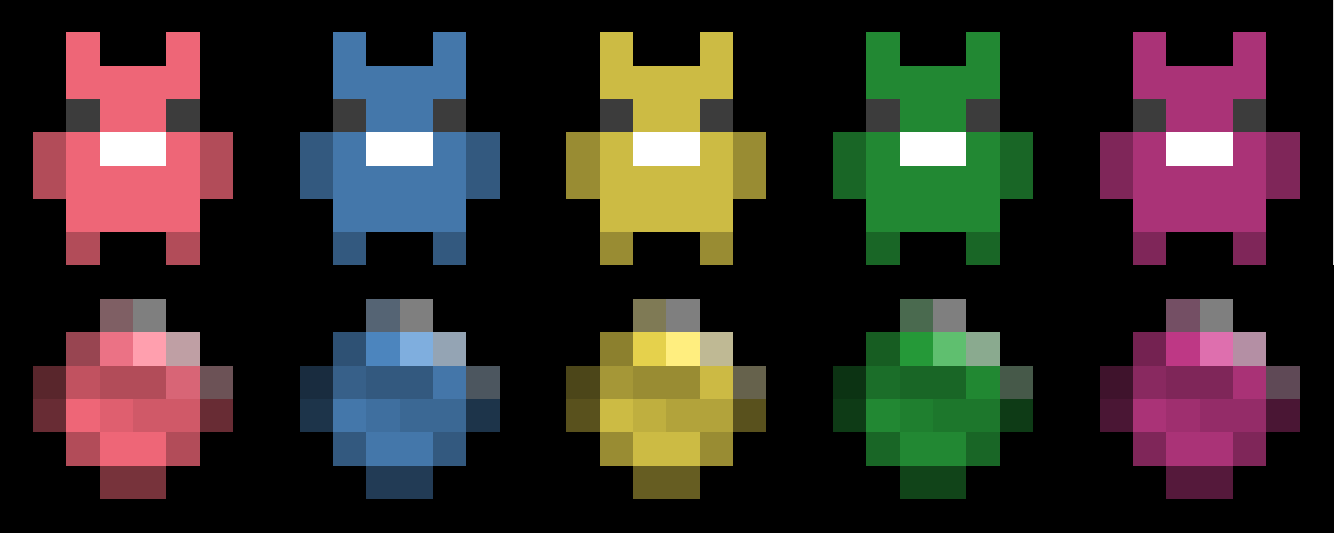} \label{fig:app/coins_entities/a}} \hspace{0.5em}
    \subfloat[]{\includegraphics[width=0.1\textwidth]{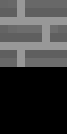} \label{fig:app/coins_entities/b}}
	\caption{Entities and sprites used in the Coins task.}
	\label{fig:app/coins_entities}
\end{figure}

The two primary entities in the Coins task are players and coins (Figure~\ref{fig:app/coins_entities/a}). We use a colorblind-friendly palette for the player and coin sprites in our experiments.

\vspace{-1em}

Each episode of Coins involves two distinct colors sampled from the five available colors (e.g., red and blue). Each player included in an episode matches one of these two colors. On each step, coins spawn on empty cells and cells occupied by players with probability $P$. When a coin spawns, it appears as one of the two colors, each with 50\% probability.

Players can enter empty cells and cells with coins in them. Each player blocks other players from entering the cell they currently occupy.

\subsubsection{Miscellaneous}

Walls (Figure~\ref{fig:app/coins_entities/b}, top row) block players from moving into a cell, and thus define the boundaries and dimensions of the room. Otherwise, players are able to freely move through empty cells (Figure~\ref{fig:app/coins_entities/b}, bottom row) in the room.

\subsection{Observations}

DeepMind Lab2D supports two different visual frames of reference~\cite{klatzky1998allocentric} for players:

\begin{enumerate}
    \item \textit{Egocentric}, meaning that players always see their avatar in an invariant position (e.g., centered) in their visual input
    \item \textit{Allocentric}, meaning that the world itself remains in an invariant position (e.g., centered) in players' visual input
\end{enumerate}

\begin{figure}[h]
	\centering
    \subfloat[Egocentric reference frame.]{\includegraphics[width=0.375\textwidth]{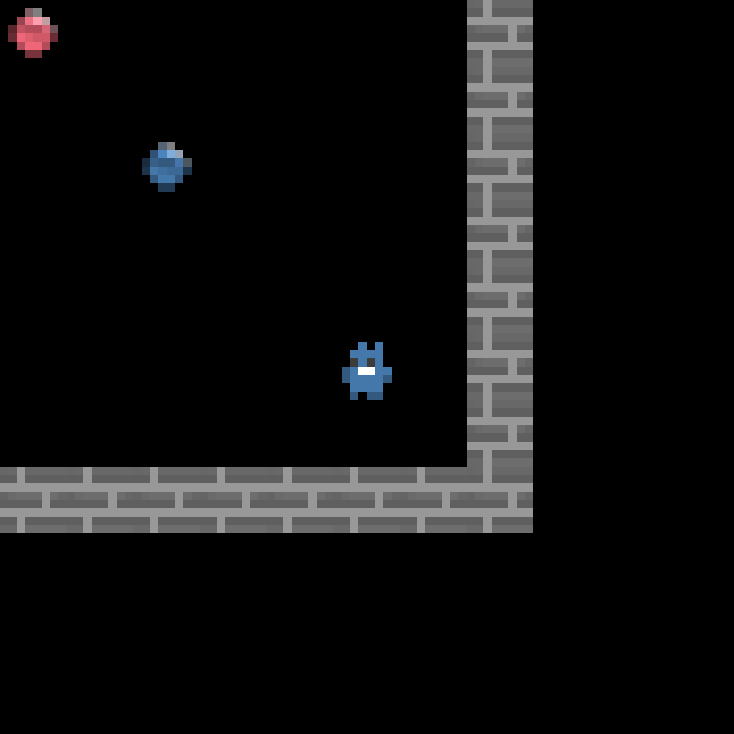} \label{fig:app/observations/a}} \hspace{1em}
    \subfloat[Allocentric reference frame.]{\includegraphics[width=0.375\textwidth]{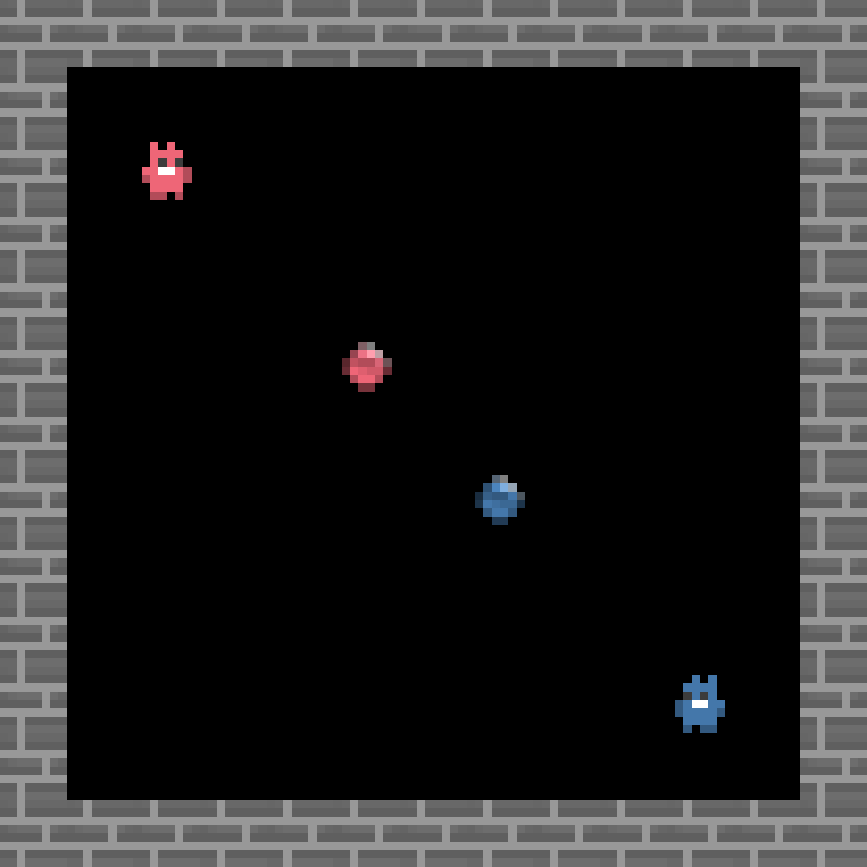} \label{fig:app/observations/b}}
	\caption{Observations for the Coins task, depicting different frames of reference for the same example game state. (a) Agent players observe Coins with an egocentric reference frame. (b) Human players observe Coins with an allocentric reference frame.}
	\label{fig:app/observations}
\end{figure}

Egocentric frames of reference promote generalization for reinforcement learning agents~\cite{hill2019environmental,ye2020rotation}. Consequently, we provide egocentric observations for our agents (Figure~\ref{fig:app/observations/a}). The agent observation spans five cells in each direction (forward, backward, left, and right) from the agent. Given the sprite dimensions ($8 \times 8 \times 3$, with the last dimension reflecting RGB channels), agents observed their environment through a $88 \times 88 \times 3$ window.

In pilot tests of our study interface, human players reported being confused by the changes in their visual observation when playing with an egocentric reference frame. In combination with prior evidence that allocentric reference frames can support human navigation in virtual environments~\cite{darken1999map}, this feedback prompted us to provide human players with an allocentric reference frame for the environment (Figure~\ref{fig:app/observations/b}). As human players move their avatar around the Coins gridworld, the room stays centered within their visual input while their avatar moves around the room.

\subsection{Actions}

Players can take one of the following five actions each step:

\begin{enumerate}
    \item \texttt{No-op}: Makes no change to the player's position.
    \item \texttt{Move up}: Moves the player up one cell.
    \item \texttt{Move down}: Moves the player down one cell.
    \item \texttt{Move left}: Moves the player left one cell.
    \item \texttt{Move right}: Moves the player right one cell.
\end{enumerate}

\clearpage
\null
\vfill

\vfill

\vfill

\clearpage
\section{Agent details} \label{sec:app/agent}

{This appendix provides detailed information on the architecture, parameterization, and training of the artificial agents that we study.}

\subsection{Agent architecture}

We build an advantage actor-critic (A2C) agent with two added algorithmic components (Figure~\ref{fig:app/agent_details}). The Social Value Orientation (SVO) component—integrated for training, evaluation, and co-play—recomputes the environment reward signal received by the agent before its use by the critic. As detailed in Section~\ref{sec:app/agent/svo}, the $\theta$ parameter guides this computation. The trembling-hand component—incorporated for evaluation and co-play—sits between the actor and the environment, replacing each action emitted by the policy with a random action with probability $\epsilon$.

\subsection{Social Value Orientation} \label{sec:app/agent/svo}

McKee et al.~\cite{mckee2020social} introduce and define the SVO algorithmic component across three descriptive levels~\cite{hamrick2020levels,marr1982philosophy}:
\begin{enumerate}
    \item At the \textit{computational level}, McKee et al. propose SVO as a mechanism ``redefin[ing] self-interest to incorporate the interests of a broader group''.
    \item At the \textit{algorithmic level}, McKee et al. introduce ``reward angles'' as a method of representing distributions of reward over the self and the group.
    \item At the \textit{implementation level}, McKee et al. offer a penalty-based approach that assigns pseudoreward to SVO agents based on the divergence between their target reward angle and the realized reward angle, in combination with a weight parameter $w$. Eq. (3) in~\cite{mckee2020social} defines this implementation.
\end{enumerate}

\begin{figure}[h]
	\centering
	\subfloat[Agent architecture. ]{\includegraphics[width=0.8\textwidth]{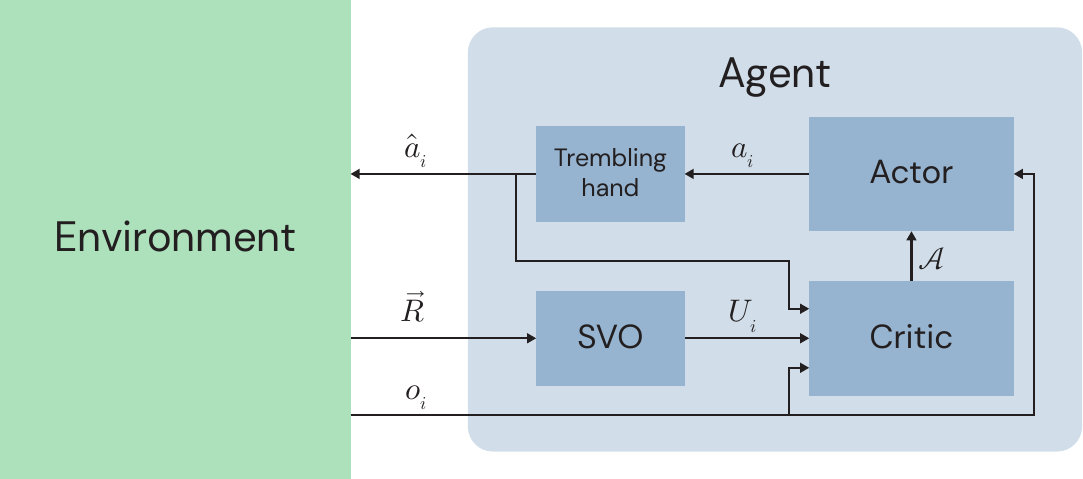}} \vspace{3em}
	\subfloat[Vector projection and SVO.]{\includegraphics[width=0.375\textwidth]{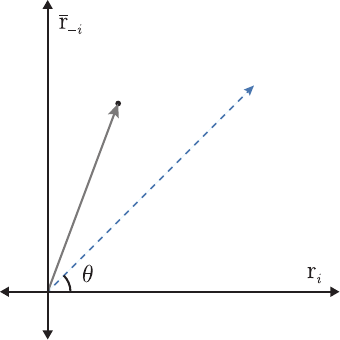} \hspace{1em} \includegraphics[width=0.375\textwidth]{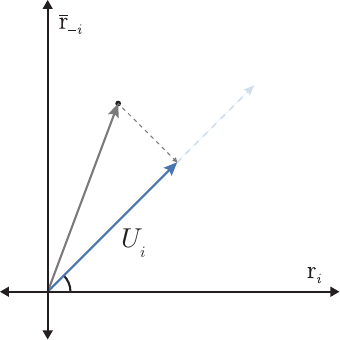}}
	\caption{Agent details. (a) In the Markov game framework, agents select actions to take on the basis of observations and rewards they receive from the environment. We construct an advantage actor-critic (A2C) agent with two algorithmic modifications: a Social Value Orientation (SVO) component that takes the group reward vector ${\vec{R}}$ as input and computes utility ${U_i}$ for the critic's use; and a ``trembling hand'' component that changes the actor's output ${a_i}$ with some probability ${\epsilon}$, producing a modified action ${\hat{a}_i}$. The agent otherwise operates as usual, with the actor initially selecting an action given the agent's observation ${o_i}$, and the critic computing an advantage estimate ${\mathcal{A}}$ from the agent's observation ${o_i}$, its utility ${U_i}$, and the agent's modified action ${\hat{a}_i}$. For more information on A2C, see~\cite{mnih2016asynchronous}. (b) The SVO component computes ${U_i}$ through vector projection. On the left, the black dot represents the reward vector given by the environment and defined by ${r_i}$ and ${\bar{r}_{-i}}$. The dotted blue line represents a vector along the agent's SVO angle. On the right, ${U_i}$ is the magnitude of the blue vector computed by projecting the environmental reward vector on the dotted blue line. See Eq.~\ref{eqn:app/vector_proj} for precise formulation.}
	\label{fig:app/agent_details}
\end{figure}

Here we introduce a new implementation of SVO that retains the computational and algorithmic levels as described above, but replaces the divergence-penalization method with an approach based on vector projection (Figure~\ref{fig:app/agent_details}). Within the Markov game framework described in~\cite{mckee2020social}, we define the vector-projection approach with a function $U_i$ to be maximized by agent $i$:

\begin{equation}
    U_{i}(s, o_i, a_i) = r_{i} \cdot \cos \theta + \bar{r}_{-i} \cdot \sin \theta
    \label{eqn:app/vector_proj}
\end{equation}

\noindent 
In the Markov game framework, $s$ is the current environmental state; $o_i$ is the observation received by agent $i$; $a_i$ is the action selected by agent $i$ through a policy $\pi(a_i|o_i)$; $r_i$ is the environmental reward received by agent $i$, contained within the state reward vector $\vec{R} = (r_1, \ldots, r_n)$ for all $n$ agents in the environment; $\theta$ is the Social Value Orientation for player $i$ (representing player $i$'s target distribution of reward among group members); and $\bar{r}_{-i}$ is a statistic summarizing the rewards of all other group members from $\vec{R}$. For these experiments, we choose to define $\bar{r}_{-i}$ as the arithmetic mean: $\bar{r}_{-i} = \frac{1}{{n - 1}} \sum_{j \neq i} r_j$, where $n$ is the group size. This formulation of SVO parallels the model of human preferences described by Griesinger and Livingston~\cite{griesinger1973toward}.

This utility function leaves most of its inputs unused, aside from information contained in the group reward vector $\vec{R}$. Consequently, the vector-projection approach may also be referenced as $U_{i}(\vec{R})$, as in Figure~\ref{fig:app/agent_details}.

\subsection{Training protocol}
Agents trained in a distributed set of environments running in parallel. Each agent trained using one GPU, for a total of $\expnumber{5}{7}$ environment steps across 150 environments. Our distributed framework checkpointed agents every 15 minutes during training.

\begin{figure}[h]
	\centering
    \includegraphics[width=\textwidth]{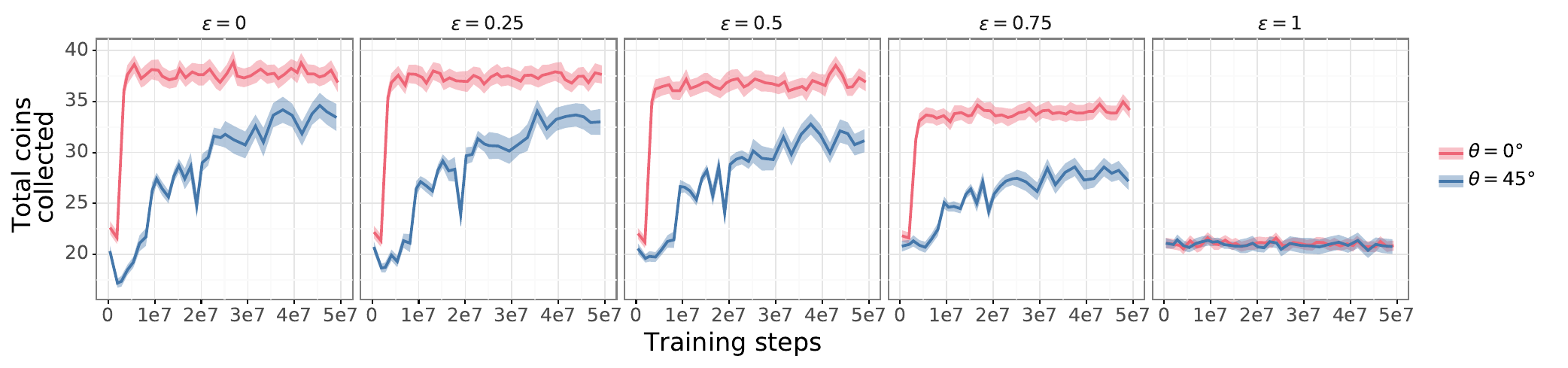}
	\caption{Effect of ${\epsilon}$ on total coin collection over training. Error bands represent 95\% confidence intervals over 100 evaluation episodes at each training checkpoint.}
	\label{fig:app/training_epsilon_all_coins}
\end{figure}

\begin{figure}[h]
	\centering
    \includegraphics[width=\textwidth]{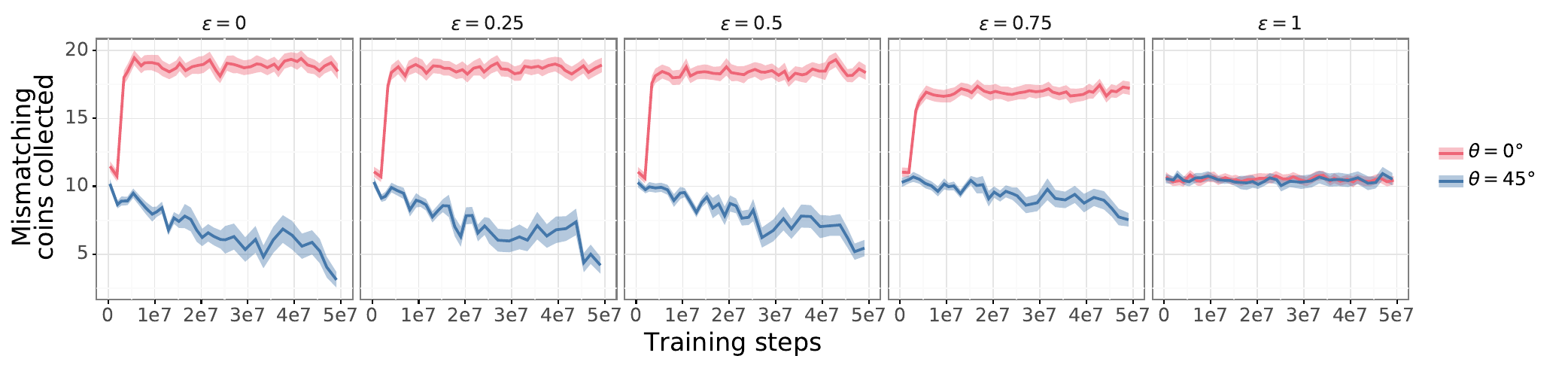}
	\caption{Effect of ${\epsilon}$ on mismatching-coin collection over training. Error bands represent 95\% confidence intervals over 100 evaluation episodes at each training checkpoint.}
	\label{fig:app/training_epsilon_mismatching_coins}
\end{figure}

\begin{figure}[h]
	\centering
    \includegraphics[width=\textwidth]{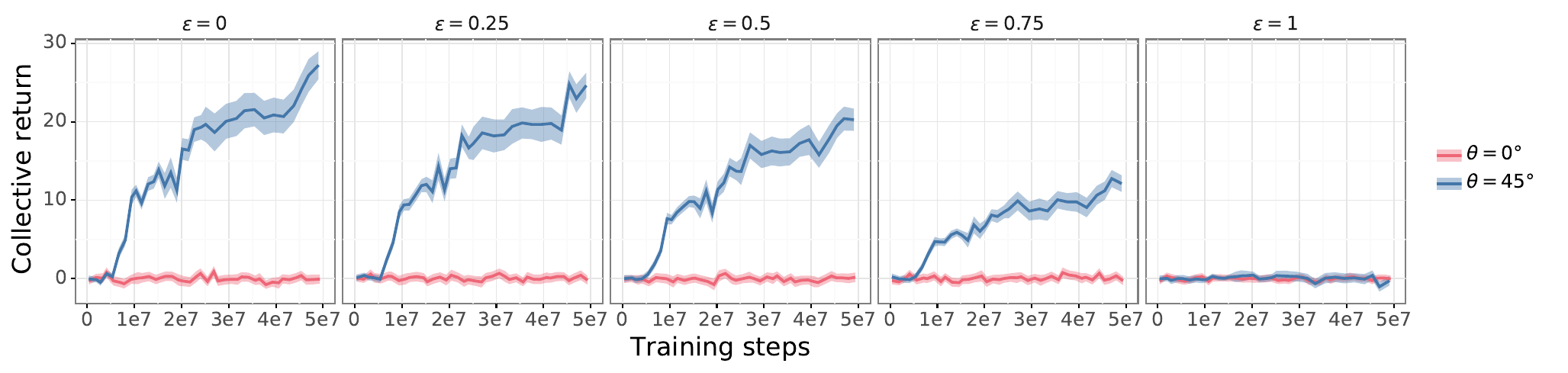}
	\caption{Effect of ${\epsilon}$ on collective return over training. Error bands represent 95\% confidence intervals over 100 evaluation episodes at each training checkpoint.}
	\label{fig:app/training_epsilon_collective_return}
\end{figure}

We compute training curves by evaluating each agent checkpoint in 100 self-play episodes, then calculating the mean and standard deviation of total coin collections (Figure~\ref{fig:app/training_epsilon_all_coins}), mismatching coin collections (Figure~\ref{fig:app/training_epsilon_mismatching_coins}), and collective return (Figure~\ref{fig:app/training_epsilon_collective_return}) for each set of evaluation episodes.

\clearpage
\vfill

\vfill

\clearpage
\null
\vfill
\vfill

\vfill

\vfill

\clearpage
\section{Human-agent interaction studies}
\label{sec:app/study}

{This appendix provides detailed information on the design of the human-agent interaction studies and on the statistical analyses conducted in each study.}

\subsection{Study details}

All three studies received a favorable opinion from the Human Behavioural Research Ethics Committee at DeepMind (\#19/004) and were approved by the Institutional Review Board for Human Subjects at Princeton University (\#11885).

Each study applied the same inclusion criteria during recruitment on Prolific: residence in the United States, prior completion of at least 20 studies on Prolific, and an approval rating of 95\% or more.

\begin{figure}[h]
    \centering
    \subfloat[Screen 1: Welcome participants to the experiment.]{\includegraphics[width=0.305\textwidth]{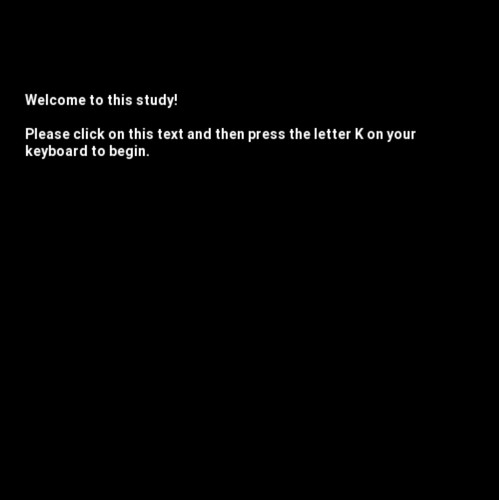}} \hspace{1em}
    \subfloat[Screen 2: Explain the keyboard controls.]{\includegraphics[width=0.305\textwidth]{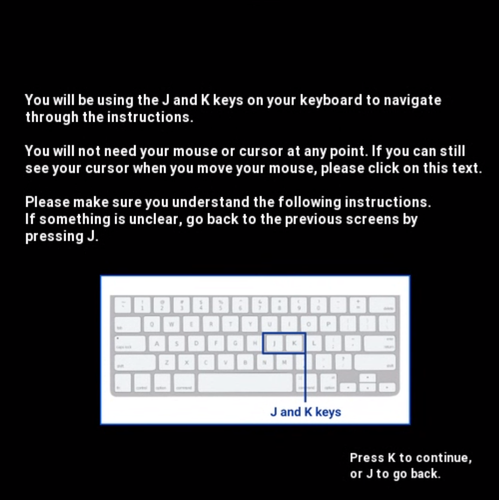}} \hspace{1em}
    \subfloat[Screen 3: Provide overview of the experiment and bonus.]{\includegraphics[width=0.305\textwidth]{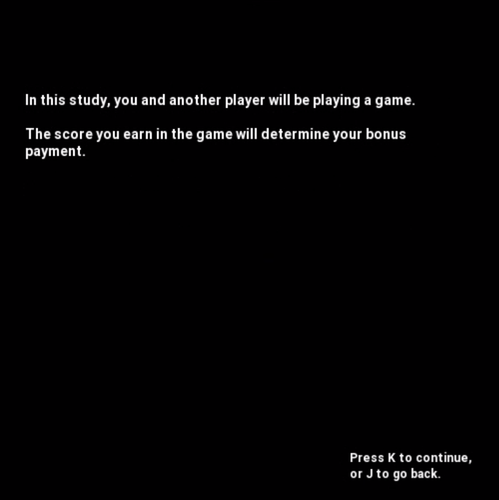}} \\
    \subfloat[Screen 4: Introduce the participant's avatar.]{\includegraphics[width=0.305\textwidth]{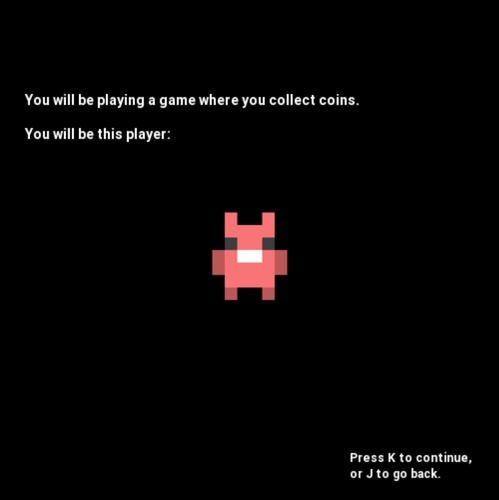}} \hspace{1em}
    \subfloat[Screen 5: Introduce the room.]{\includegraphics[width=0.305\textwidth]{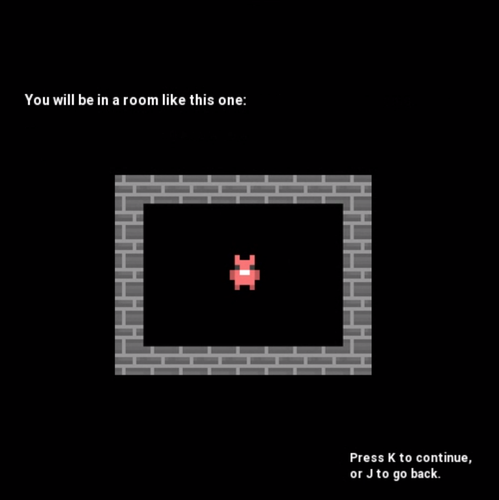}} \hspace{1em}
    \subfloat[Screen 6: Explain the controls for movement.]{\includegraphics[width=0.305\textwidth]{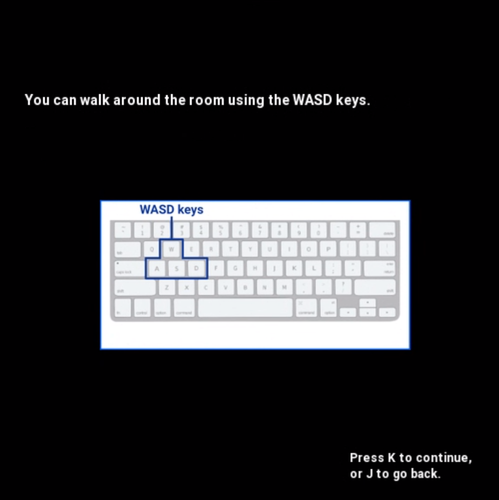}} \\
    \subfloat[Screen 7: Explain coin collections.]{\includegraphics[width=0.305\textwidth]{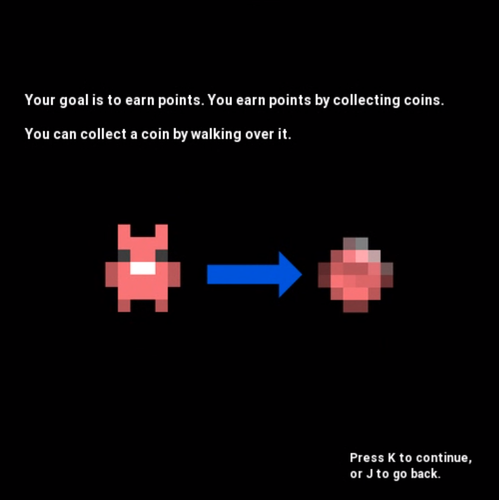}} \hspace{1em}
    \subfloat[Screen 8: Introduce the practice episode.]{\includegraphics[width=0.305\textwidth]{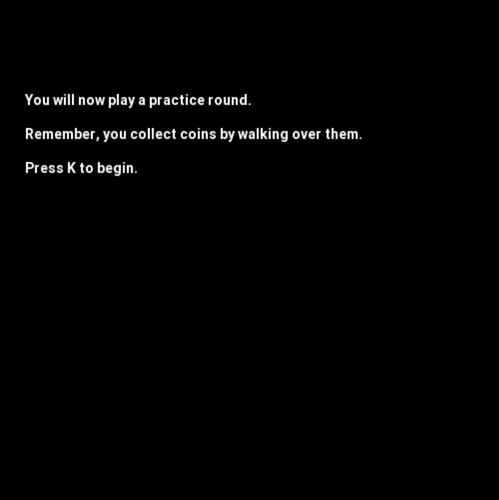}}
    \caption{Screenshots of instruction screens in Study 1.}
    \label{fig:app/screenshots_1}
\end{figure}

\begin{figure}[h]
    \subfloat[Screen 9: Load episode.]{\includegraphics[width=0.305\textwidth]{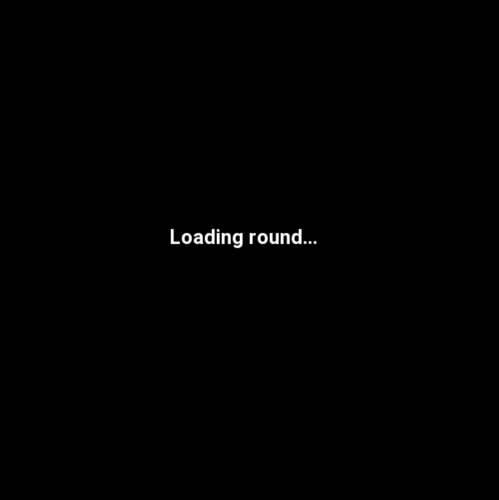}} \hspace{1em}
    \subfloat[Screen 10: Play practice episode.]{\includegraphics[width=0.305\textwidth]{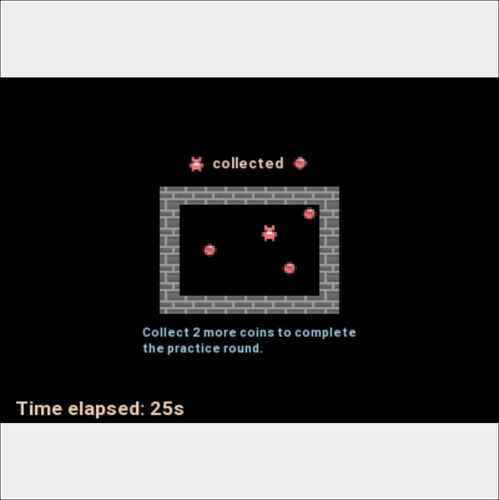}} \hspace{1em}
    \subfloat[Screen 11: Save episode data.]{\includegraphics[width=0.305\textwidth]{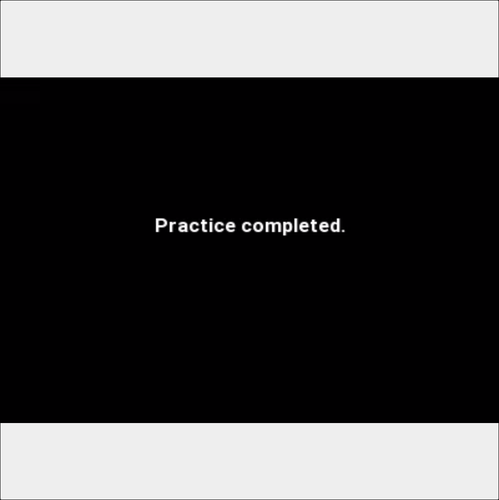}}
    \caption{Screenshots of practice episode in Study 1.}
    \label{fig:app/screenshots_2}
\end{figure}

\begin{figure}[h]
    \centering
    \subfloat[Screen 12: Explain the bonus rules.]{\includegraphics[width=0.305\textwidth]{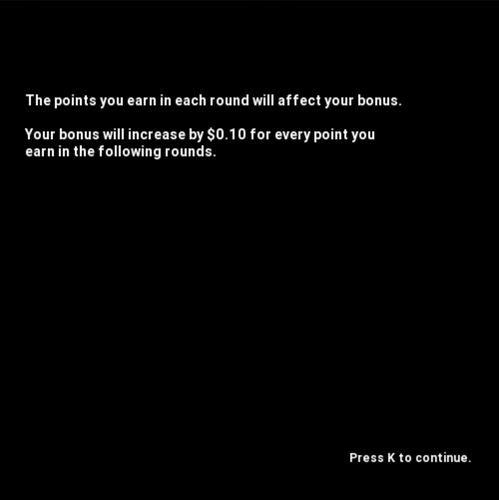} \label{fig:app/screenshots_3/a}} \hspace{1em}
    \subfloat[Screen 13: Explain co-play.]{\includegraphics[width=0.305\textwidth]{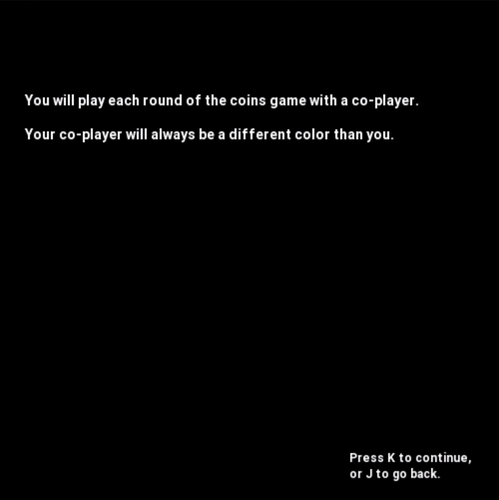}} \hspace{1em}
    \subfloat[Screen 14: Introduce an example co-player.]{\includegraphics[width=0.305\textwidth]{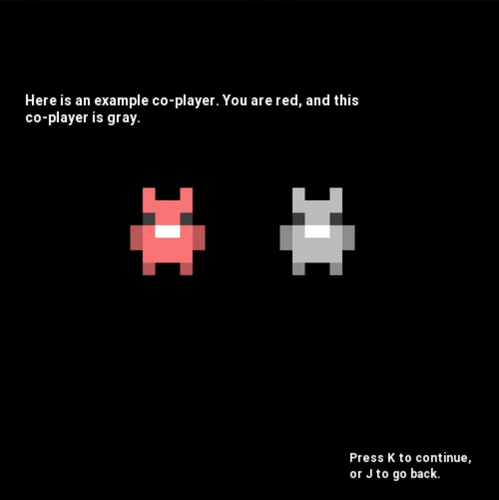}} \hspace{1em}
    \caption{Screenshots of instruction screens in Study 1.}
    \label{fig:app/screenshots_3}
\end{figure}

\begin{figure}[h]
    \centering
    \subfloat[Screen 15: Explain coin colors.]{\includegraphics[width=0.305\textwidth]{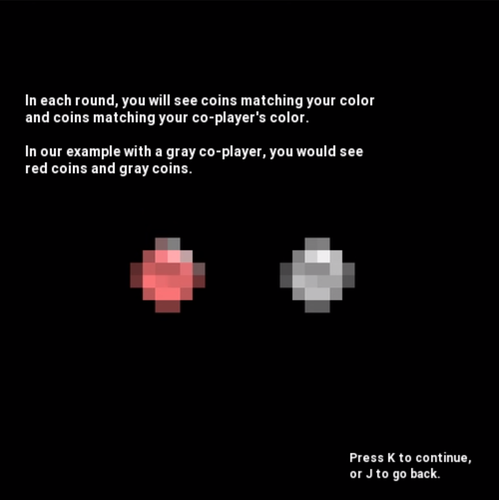}} \hspace{1em}
    \subfloat[Screen 16: Explain matching coin rules.]{\includegraphics[width=0.305\textwidth]{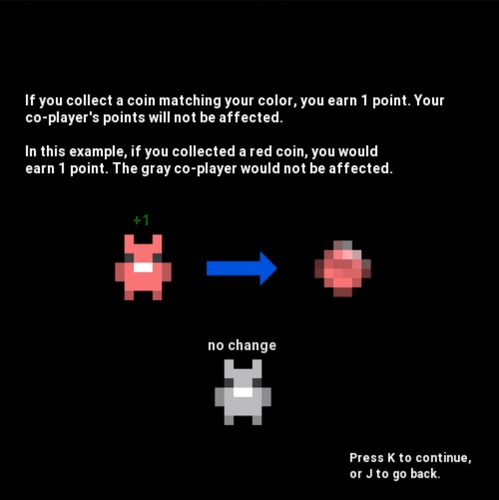} \label{fig:app/screenshots_4/b}} \hspace{1em}
    \subfloat[Screen 17: Explain matching coin rules.]{\includegraphics[width=0.305\textwidth]{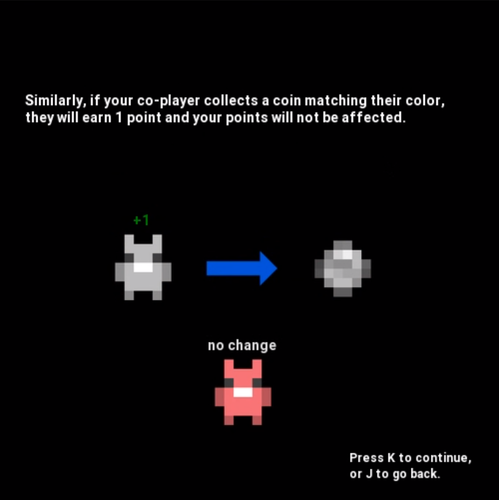}} \\
    \subfloat[Screen 18: Explain mismatching coin rules.]{\includegraphics[width=0.305\textwidth]{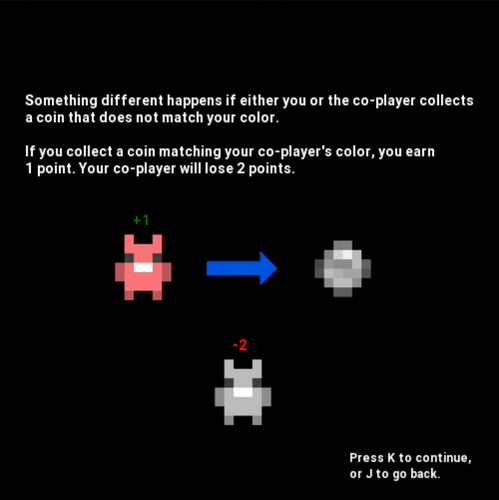}} \hspace{1em}
    \subfloat[Screen 19: Explain mismatching coin rules.]{\includegraphics[width=0.305\textwidth]{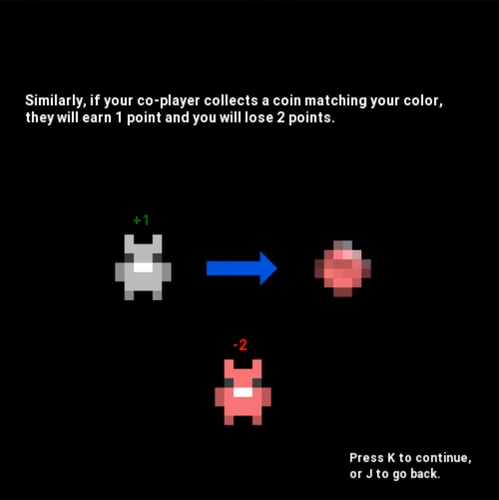} \label{fig:app/screenshots_4/e}} \hspace{1em}
    \subfloat[Screen 20: Introduce all co-players.]{\includegraphics[width=0.305\textwidth]{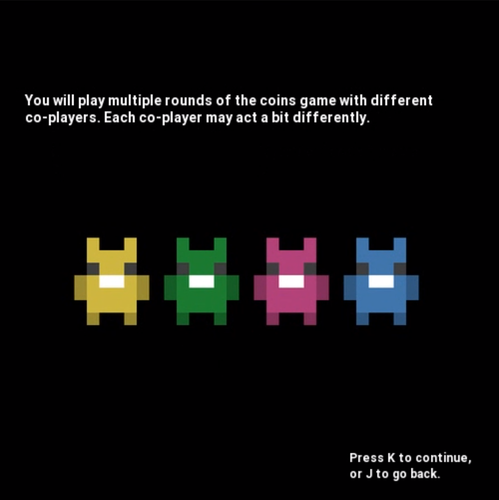}} \\
    \subfloat[Screen 21: Explain questions.]{\includegraphics[width=0.305\textwidth]{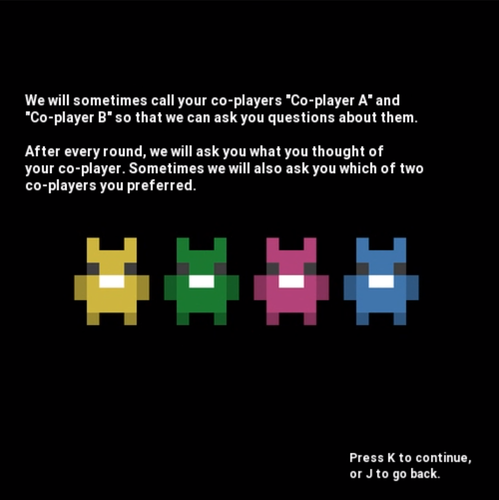}} \hspace{1em}
    \subfloat[Screen 22: Begin co-play episodes.]{\includegraphics[width=0.305\textwidth]{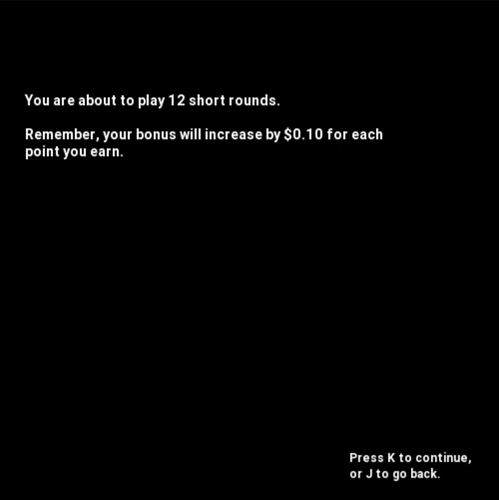} \label{fig:app/screenshots_4/h}}
    \caption{Screenshots of instruction screens in Study 1.}
    \label{fig:app/screenshots_4}
\end{figure}

\begin{figure}[h]
    \centering
    \subfloat[Screen 23: Introduce co-player for this episode.]{\includegraphics[width=0.305\textwidth]{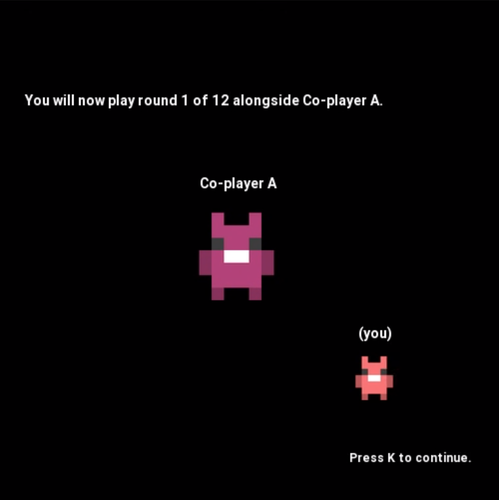}} \hspace{1em}
    \subfloat[Screen 24: Load episode.]{\includegraphics[width=0.305\textwidth]{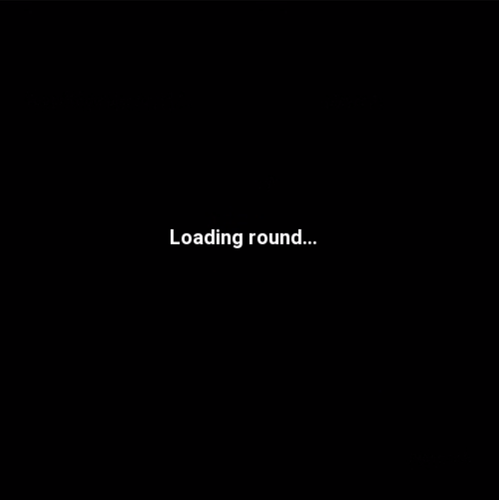}} \hspace{1em}
    \subfloat[Screen 25: Play co-play episode.]{\includegraphics[width=0.305\textwidth]{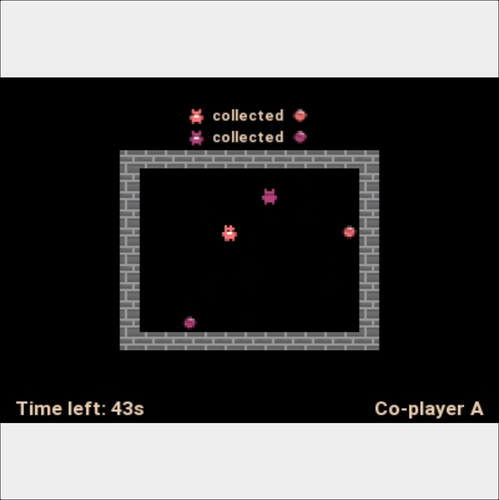}} \\
    \subfloat[Screen 26: Save episode data.]{\includegraphics[width=0.305\textwidth]{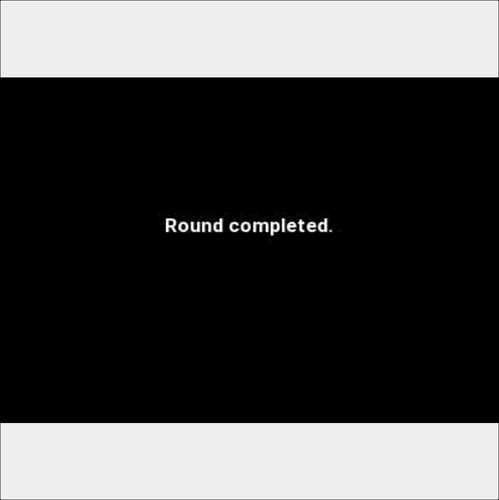}} \hspace{1em}
    \subfloat[Screen 27: Elicit judgment.]{\includegraphics[width=0.305\textwidth]{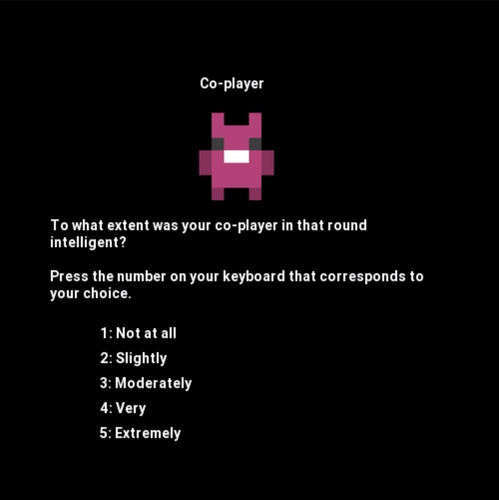}} \hspace{1em}
    \subfloat[Screen 28: Confirm judgment.]{\includegraphics[width=0.305\textwidth]{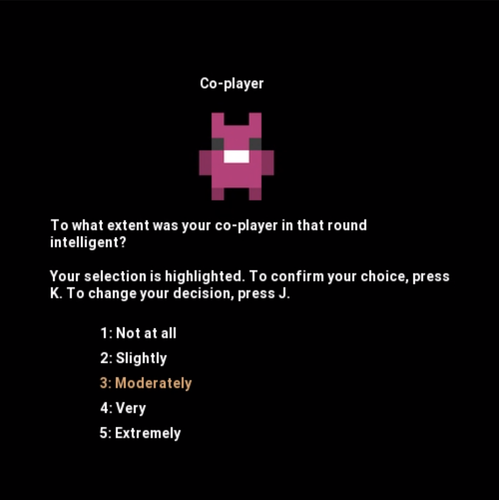}}
    \caption{Screenshots of an ``A'' co-play episode and subsequent questions on perceptions in Study 1. {(e) and (f) show the screens for eliciting and confirming one judgment; the study repeats these screens for all four judgments, randomizing the sequence of judgments for each episode.}}
    \label{fig:app/screenshots_5}
\end{figure}

\begin{figure}[h]
    \centering
    \subfloat[Screen 29: Introduce co-player for this episode.]{\includegraphics[width=0.305\textwidth]{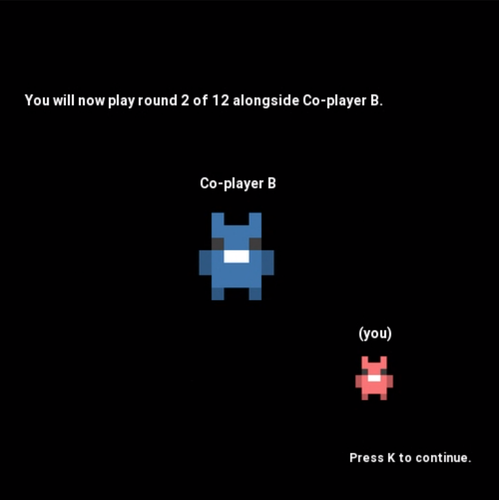}} \hspace{1em}
    \subfloat[Screen 30: Load episode.]{\includegraphics[width=0.305\textwidth]{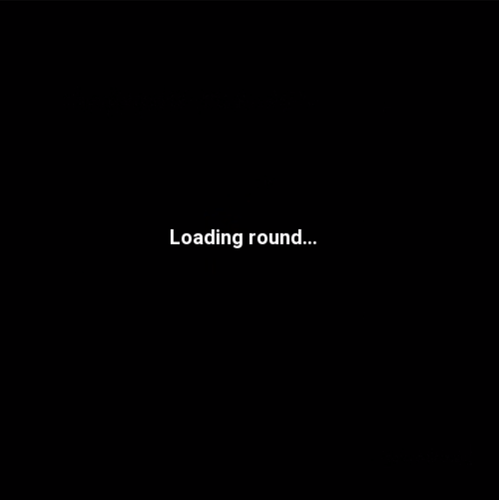}} \hspace{1em}
    \subfloat[Screen 31: Play co-play episode.]{\includegraphics[width=0.305\textwidth]{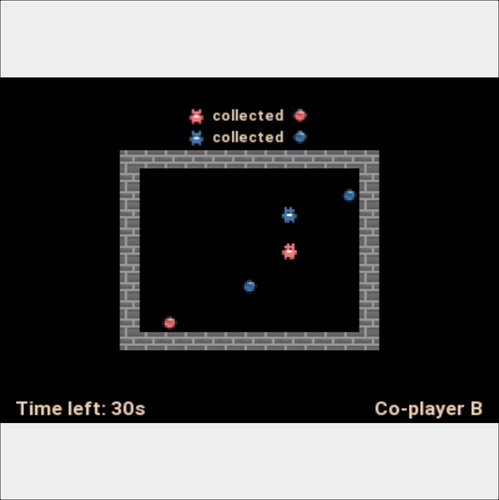}} \\
    \subfloat[Screen 32: Save episode data.]{\includegraphics[width=0.305\textwidth]{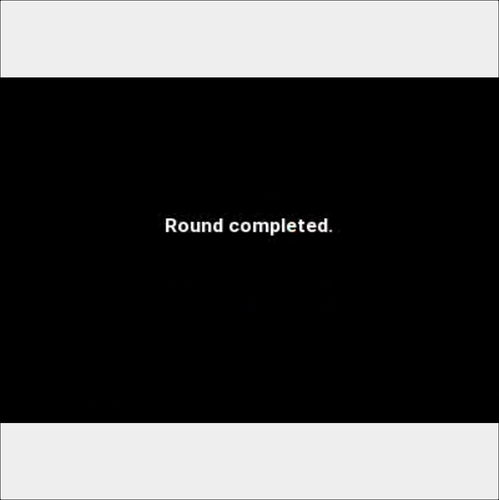}} \hspace{1em}
    \subfloat[Screen 33: Elicit judgment.]{\includegraphics[width=0.305\textwidth]{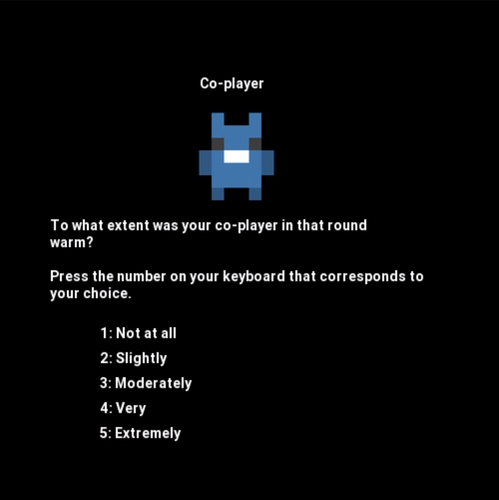}} \hspace{1em}
    \subfloat[Screen 34: Confirm judgment.]{\includegraphics[width=0.305\textwidth]{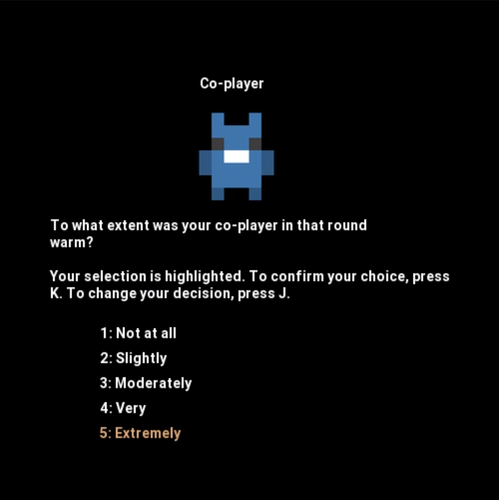}} \\
    \subfloat[Screen 35: Elicit preference.]{\includegraphics[width=0.305\textwidth]{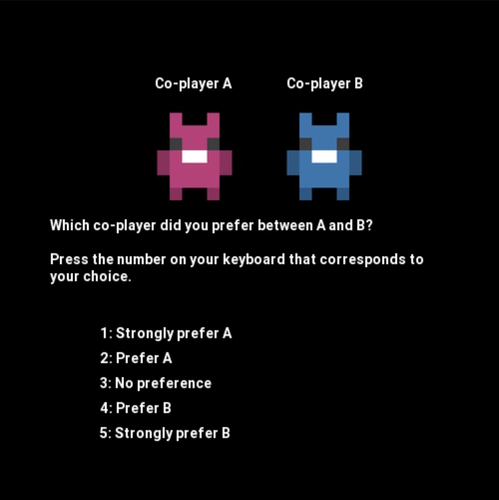}} \hspace{1em}
    \subfloat[Screen 36: Confirm preference.]{\includegraphics[width=0.305\textwidth]{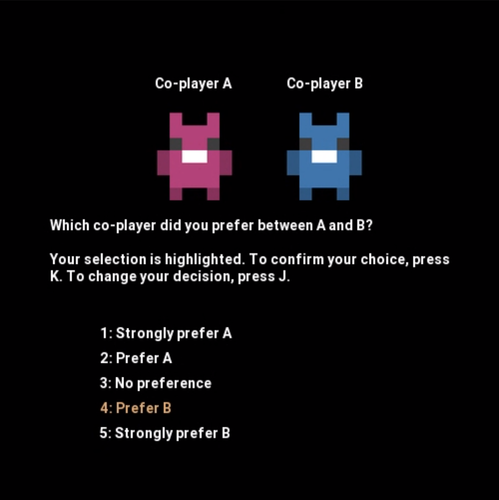}}
    \caption{Screenshots of a ``B'' co-play episode and subsequent questions on perceptions and preferences in Study 1. {(e) and (f) show the screens for eliciting and confirming one judgment; the study repeats these screens for all four judgments, randomizing the sequence of judgments for each episode.}}
    \label{fig:app/screenshots_6}
\end{figure}

\begin{figure}[h]
    \centering
    \subfloat[Screen 37: Confirm bonus and transition to post-task questionnaire.]{\includegraphics[width=0.305\textwidth]{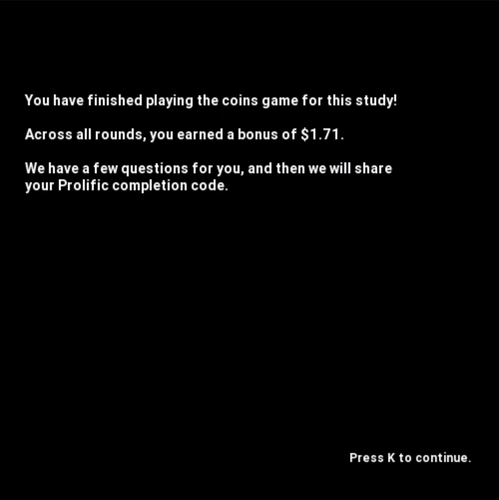}}
    \caption{Screenshot of pre-questionnaire page in Study 1.}
    \label{fig:app/screenshots_7}
\end{figure}

We include screenshots showing how Study 1 unfolded for each participant:

\begin{enumerate}
    \item Read general instructions on the study and gameplay (Figure~\ref{fig:app/screenshots_1}).
    \item Play practice episode (Figure~\ref{fig:app/screenshots_2}).
    \item Read instructions on game rules and co-players (Figures~\ref{fig:app/screenshots_3} and~\ref{fig:app/screenshots_4}).
    \item Play co-play episode with Co-player A and answer questions about perceptions (Figure~\ref{fig:app/screenshots_5}).
    \item Play co-play episode with Co-player B and answer questions about perceptions and preferences (Figure~\ref{fig:app/screenshots_6}).
    \item Repeat steps 3 and 4 for another 10 episodes.
    \item Transition to post-task questionnaire (Figure~\ref{fig:app/screenshots_7}).
\end{enumerate}

\begin{figure}[h]
    \centering
    \includegraphics[width=0.62\textwidth]{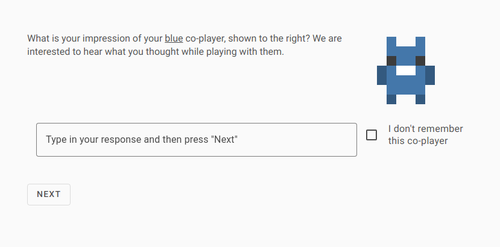}
    \caption{Screenshot of open-ended impression question from post-task questionnaire in Study 1.}
    \label{fig:app/open_ended_impression_question}
\end{figure}

Figure~\ref{fig:app/open_ended_impression_question} shows an example open-ended impression question from the post-task questionnaire.

During gameplay, a ``ticker'' at the top of the screen displayed the three most recent coin collection events. The study interface did not provide participants with cumulative metrics (e.g., score or coin collections).

In all three studies, the practice episode instructed participants to collect five coins. The practice episode ended after the participant collected five coins or five minutes elapsed, whichever occurred first.

In Studies 1 and 2, we generated the sequence of co-players for each participant by listing the 12 possible pairwise combinations of agents, randomly shuffling the order of the pairs, and then randomly shuffling the order of the agents within each pair.

Study 2 closely resembled Study 1, aside from identifying the bonus earned per point as $\$0.02$ (Figures~\ref{fig:app/screenshots_3/a} and~\ref{fig:app/screenshots_4/h}) and using the reward values from the shifted incentive structure (Table~\ref{tab:payoff_table_offset}) when explaining the rules for matching and mismatching coin collections (Figures~\ref{fig:app/screenshots_4/b}-\ref{fig:app/screenshots_4/e}).

\begin{figure}[h]
    \centering
    \subfloat[Screen 20: Explain questions.]{\includegraphics[width=0.305\textwidth]{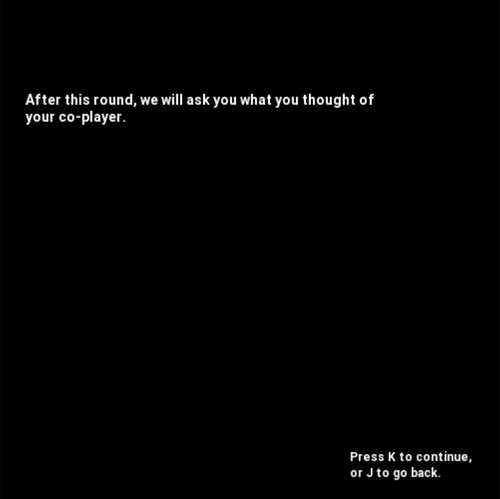}} \hspace{1em}
    \subfloat[Screen 21: Begin co-play episodes.]{\includegraphics[width=0.305\textwidth]{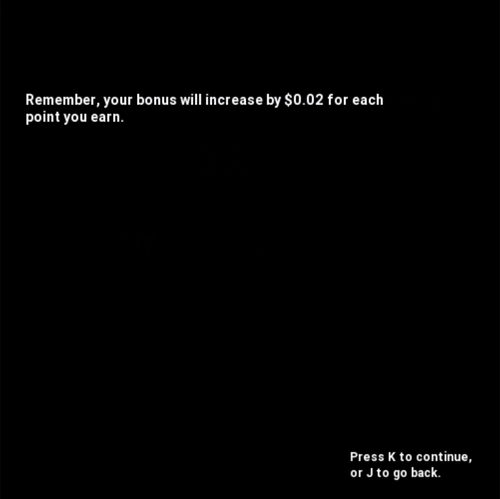}} \hspace{1em}
    \subfloat[Screen 22: Introduce co-player for this episode.]{\includegraphics[width=0.305\textwidth]{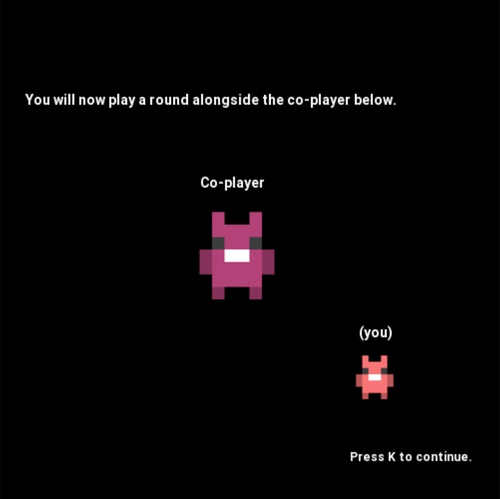}} \\
    \subfloat[Screen 23: Load episode.]{\includegraphics[width=0.305\textwidth]{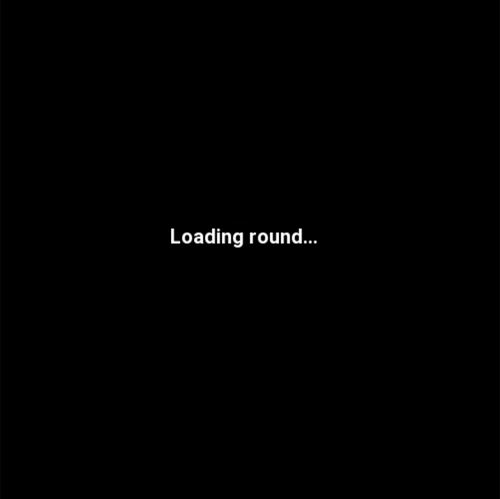}} \hspace{1em}
    \subfloat[Screen 24: Play co-play episode.]{\includegraphics[width=0.305\textwidth]{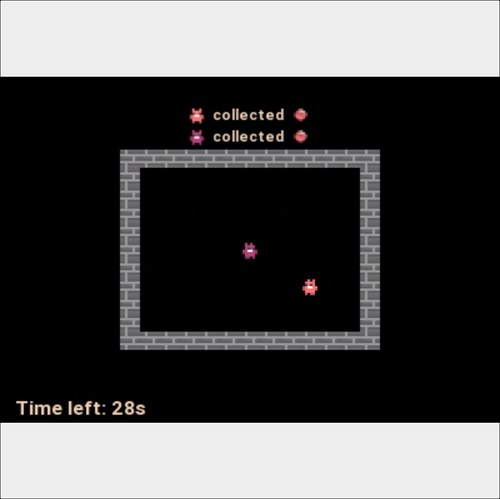}} \hspace{1em}
    \subfloat[Screen 25: Save episode data.]{\includegraphics[width=0.305\textwidth]{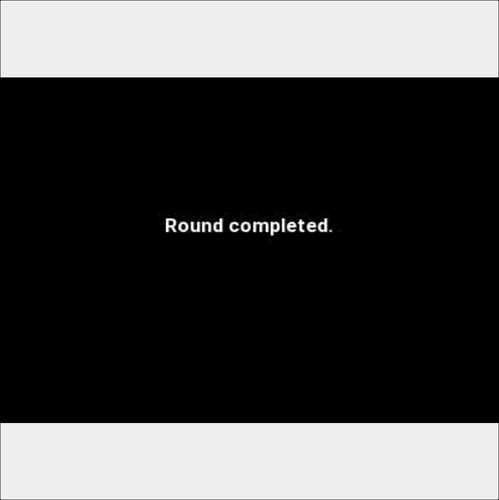}} \\
    \subfloat[Screen 26: Elicit judgment.]{\includegraphics[width=0.305\textwidth]{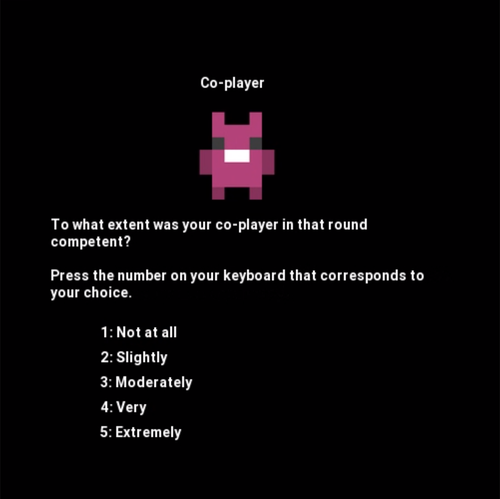}} \hspace{1em}
    \subfloat[Screen 27: Confirm judgment.]{\includegraphics[width=0.305\textwidth]{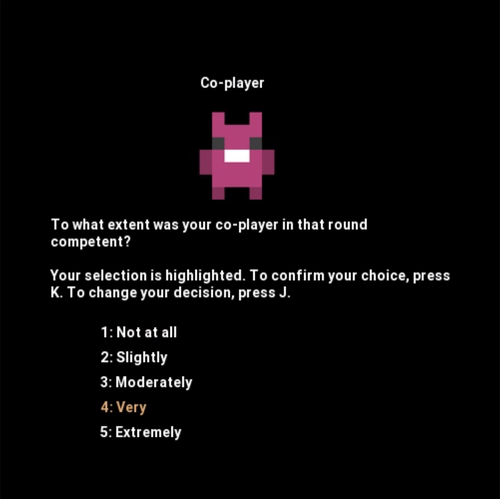}}
    \caption{Screenshots of instructions, co-play episode, and subsequent questions on perceptions in Study 3. {(g) and (h) show the screens for eliciting and confirming one judgment; the study repeats these screens for all four judgments, randomizing the sequence of judgments for each episode.}}
    \label{fig:app/screenshots_8}
\end{figure}

\begin{figure}[h]
    \centering
    \subfloat[Screen 28: Introduce partner-choice episode.]{\includegraphics[width=0.305\textwidth]{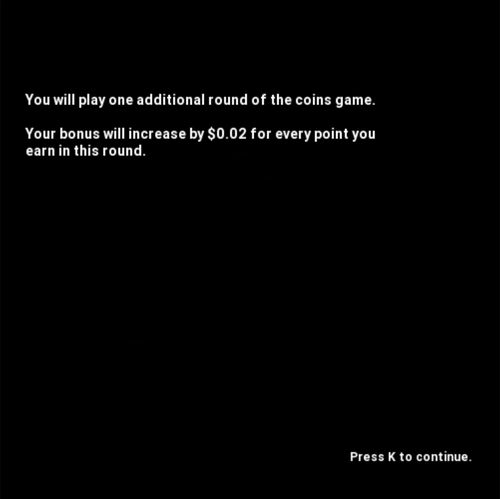}} \hspace{1em}
    \subfloat[Screen 29: Explain partner choice.]{\includegraphics[width=0.305\textwidth]{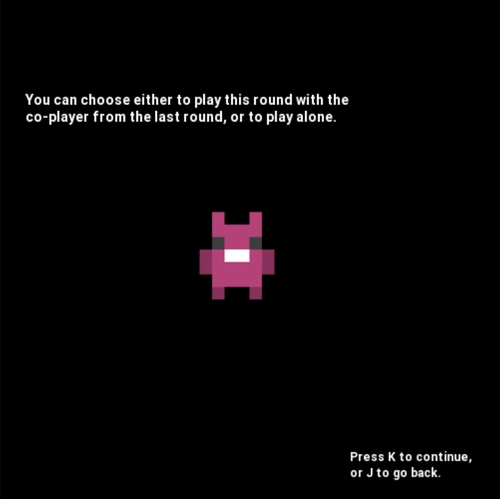}} \hspace{1em}
    \subfloat[Screen 30: Explain coins in partner-choice episode.]{\includegraphics[width=0.305\textwidth]{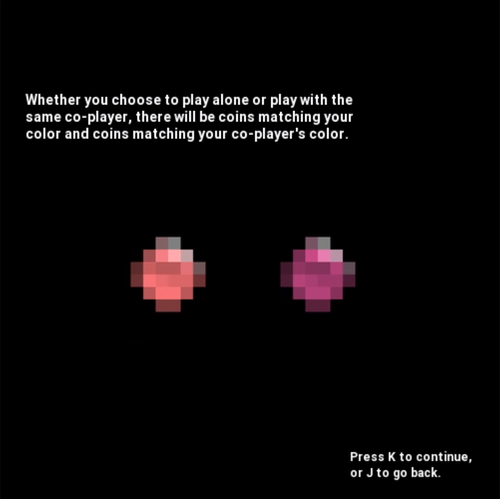}} \\
    \subfloat[Screen 31: Elicit partner choice.]{\includegraphics[width=0.305\textwidth]{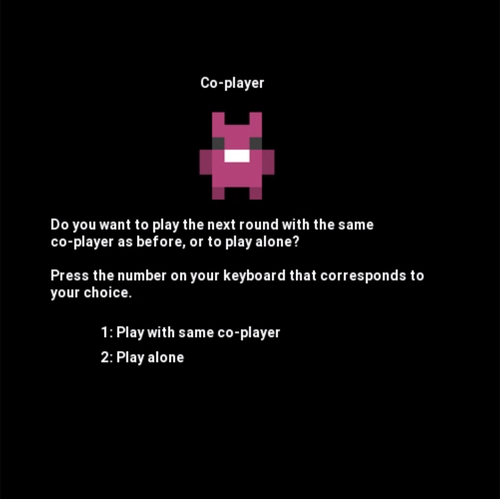}} \hspace{1em}
    \subfloat[Screen 32: Confirm partner choice.]{\includegraphics[width=0.305\textwidth]{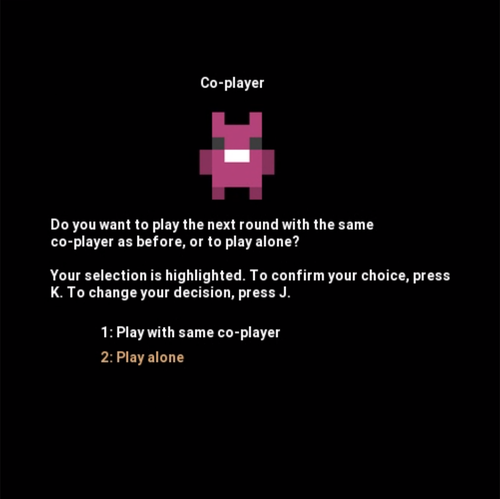}} \hspace{1em}
    \subfloat[Screen 33: Introduce players for this episode.]{\includegraphics[width=0.305\textwidth]{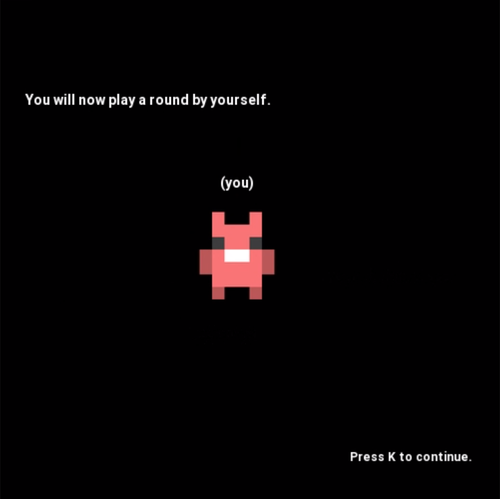}} \\
    \subfloat[Screen 34: Load episode.]{\includegraphics[width=0.305\textwidth]{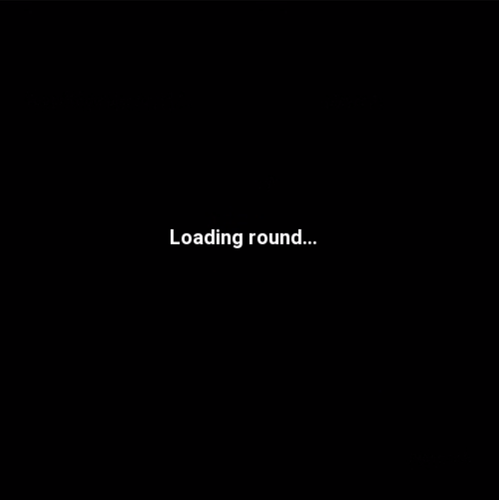}} \hspace{1em}
    \subfloat[Screen 35: Play partner-choice episode.]{\includegraphics[width=0.305\textwidth]{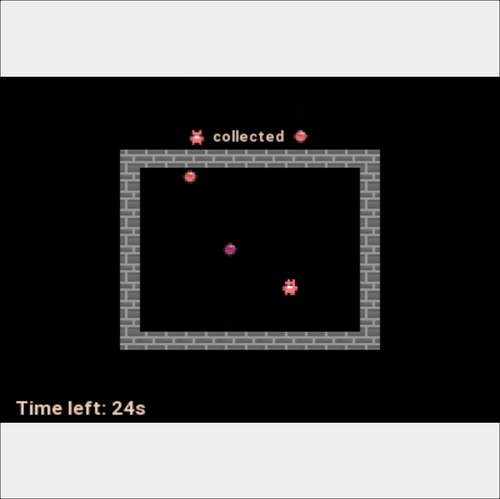}} \hspace{1em}
    \subfloat[Screen 35: Save episode data.]{\includegraphics[width=0.305\textwidth]{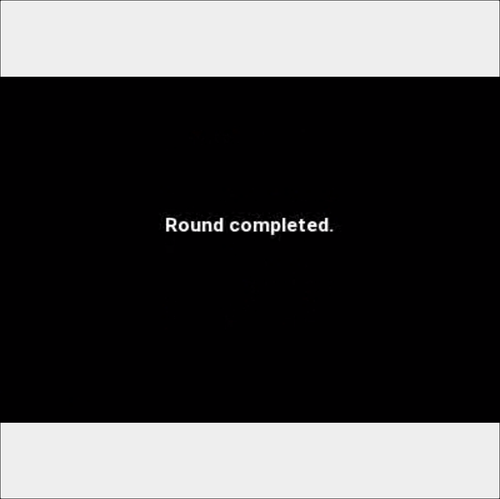}}
    \caption{Screenshots of instructions, partner choice question, and partner-choice episode in Study 3.}
    \label{fig:app/screenshots_9}
\end{figure}

Study 3 largely preserved the instructions preceding the first co-play episode (Figures~\ref{fig:app/screenshots_1}-\ref{fig:app/screenshots_4}), though the wording on certain screens changed slightly to refer to a single upcoming episode rather than multiple episodes (e.g., ``the next round'' rather than ``each round''). The study branched more substantially starting just before the first co-play episode, as shown in Figures~\ref{fig:app/screenshots_8} and~\ref{fig:app/screenshots_9}. After playing a single co-play episode, participants read a description of their partner choice for the following episode. They played the partner-choice episode as they chose (that is, either alone or with the co-player from the previous episode).

\subsection{Analytic details}

For our statistical analyses, we rely on several regression methods to help examine our data. {In situations where our data consist of independent observations, we employ standard linear and logistic regressions to analyze the relationships between variables. For analyses involving repeated measurements collected from each participant, we leverage generalized linear mixed-effect models (GLMMs). These models offer flexibility to model different types of outcome measures (e.g., preferences reported on a bounded scale) and account for non-independent observations.}

In addition to fitting regression models, we leverage the ANOVA method (standing for ``analysis of variance''). ANOVAs~\cite{fisher1928statistical} allow us to test whether changing the value of an independent variable (e.g., an agent hyperparameter) significantly affects the value of a specified dependent variable (e.g., human perceptions of warmth or competence). Each ANOVA is summarized with an $F$-statistic, two subsetted values on the $F$-statistic (the between-groups degrees of freedom and the within-groups degrees of freedom), and a $p$-value.

\subsubsection{Study 1}

We expect the social perception questions to exhibit suitable psychometric properties: namely, participants should offer relatively consistent judgments across repeated interactions with a given agent; participants should vary more in their judgments between traits than within a particular trait; and the two items within each composite measure (``warm'' and ``well-intentioned'' for warmth, ``competent'' and ``intelligent'' for competence) should correlate consistently.

As reported in the main text (Table~\ref{tab:study_1_consistency}), participants made highly consistent judgments for a given agent on a given trait. A mixed ANOVA indicated that social perceptions varied more between traits than within a particular trait, $F_{4,3744} = 96.2$, $p < 0.001$. Finally, both the composite warmth measure ($\rho = 0.93$) and the composite competence measure ($\rho = 0.92$) exhibit high internal consistency, as measured by the Spearman-Brown formula.

\begin{figure}[h]
	\centering
    \subfloat[Warmth.]{\includegraphics[height=4.5cm]{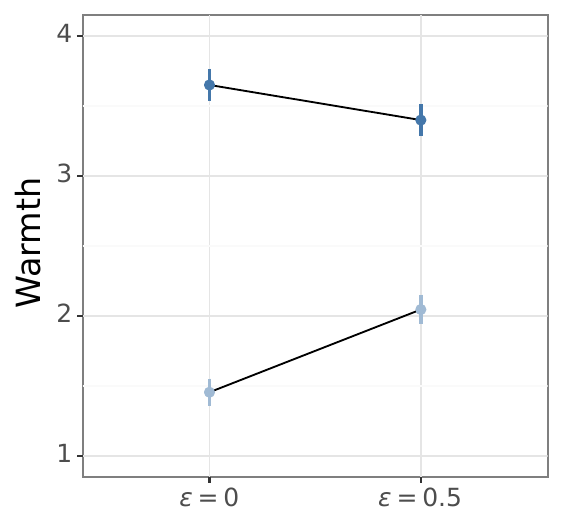} \label{fig:app/study_1_interaction_effects/a}} \hspace{1em}
    \subfloat[Competence.]{\includegraphics[height=4.5cm]{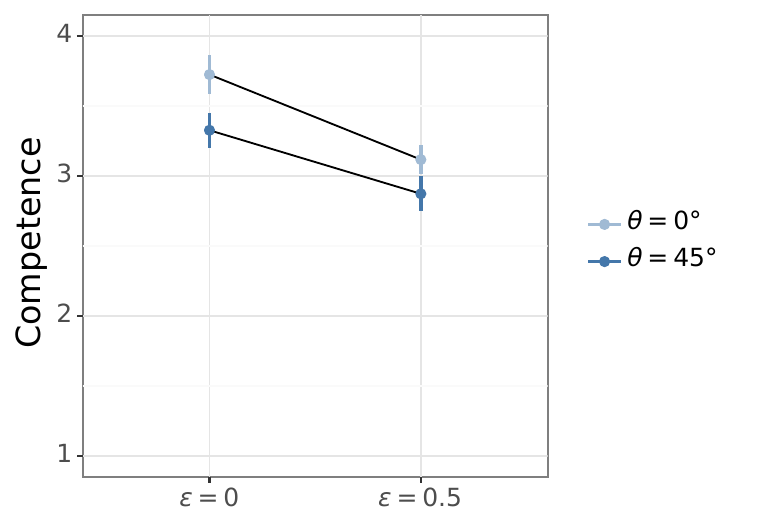} \label{fig:app/study_1_interaction_effects/b}}
    \captionsetup{width=.75\textwidth}
	\caption{Interaction between the SVO and trembling-hand components on social perceptions in Study 1. Error bars reflect 95\% confidence intervals.}
	\label{fig:app/study_1_interaction_effects}
\end{figure}

A two-way mixed ANOVA modeled the effects of $\theta$, $\epsilon$, and their interaction on warmth evaluations. Figure~\ref{fig:study_1_main_effects/a} depicts the main effect of $\theta$ on perceived warmth (that is, marginalized over $\epsilon$ values), and Figure~\ref{fig:app/study_1_interaction_effects/a} visualizes the full two-way interaction. The following means and standard deviations help describe main effects by marginalizing over all other variables. Participants perceived $\theta = 45\degree$ agents ($m = 3.52$, $sd = 1.40$) as significantly warmer than $\theta = 0\degree$ agents ($m = 1.75$, $sd = 1.31$), $F_{1,1108} = 1006.8$, $p < 0.001$. The $\epsilon$ parameter also exerted a significant effect on warmth judgments, with $\epsilon = 0.5$ agents ($m = 2.73$, $sd = 1.33$) perceived as significantly warmer than $\epsilon = 0$ agents ($m = 2.55$, $sd = 1.49$), $F_{1,1108} = 9.2$, $p = 0.002$. Finally, the effect of the interaction between the $\theta$ and $\epsilon$ parameters was significant, $F_{1,1108} = 56.7$, $p < 0.001$.

A two-way mixed ANOVA modeled the effects of $\theta$, $\epsilon$, and their interaction on competence judgments. Figure~\ref{fig:study_1_main_effects/b} depicts the main effect of $\epsilon$ on competence evaluations (that is, marginalized over $\theta$ values), and Figure~\ref{fig:app/study_1_interaction_effects/b} visualizes the full two-way interaction. The following means and standard deviations help describe main effects by marginalizing over all other variables. Participants judged $\epsilon = 0$ agents ($m = 3.53$, $sd = 1.28$) as significantly more competent than $\epsilon = 0.5$ agents ($m = 3.00$, $sd = 1.18$), $F_{1,1108} = 70.6$, $p < 0.001$. The $\theta$ parameter also exerted a significant effect on competence evaluations, $F_{1,1108} = 25.8$, $p < 0.001$. Participants perceived $\theta = 0\degree$ agents ($m = 3.42$, $sd = 1.29$) as significantly more competent than $\theta = 45\degree$ agents ($m = 3.10$, $sd = 1.21$). The effect of the interaction between the $\theta$ and $\epsilon$ parameters was not significant, $F_{1,1108} = 1.5$, $p = 0.22$.

A sequence of fractional response models evaluated the predictive value of the hypothesized predictors for participants' stated pairwise preferences. We construct each model through a GLMM with a binomial link function, normalizing the full preference scale with zero and one as its endpoints. To ensure model comparability, all fractional response models predict the exact same outcome variable.

\begin{figure}[h]
	\centering
    \includegraphics[height=4.75cm]{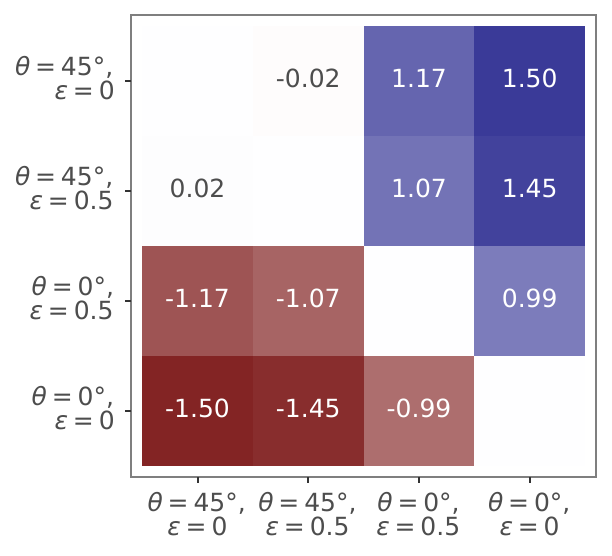}
    \captionsetup{width=.75\textwidth}
	\caption{Pairwise preferences for different agents in Study 1. Cell values represent self-reported preferences for the row agent over the column agent.}
	\label{fig:app/study_1_preference_matrix}
\end{figure}

The first GLMM predicted stated preferences from a joint intercept, a fixed effect for the identities of the agents being compared, and a random effect for the participant.
The underlying identities of the agents exerted a significant influence on the participant's stated pairwise preferences, $\chi^2(11) = 572.8$, $p < 0.001$. Figure~\ref{fig:app/study_1_preference_matrix} shows mean pairwise preferences calculated for each pair of agent identities. 
The effect of the joint intercept was not significant, $p = 0.32$.
The model achieves an AIC of $1661.9$ and $R^2_m = 0.362$.

The second GLMM predicted stated preferences from a joint intercept, a fixed effect for the difference in scores the participant received when playing with each of the agents being compared, and a random effect for the participant.
The difference in scores earned by the participant significantly affected stated pairwise preferences, $\mathrm{OR} = 1.12$, 95\% CI $[1.11, 1.13]$, $p < 0.001$. 
The effect of the joint intercept was not significant, $p = 0.12$.
The model achieves an AIC of $1697.2$ and $R^2_m = 0.363$.

The third GLMM predicted stated preferences from a joint intercept, a fixed effect for the difference in perceived warmth for each of the agents being compared, a fixed effect for the difference in perceived competence for each of the agents, and a random effect for the participant.
The difference in the perceived warmth of the agents significantly influenced stated pairwise preferences, $\mathrm{OR} = 2.23$, 95\% CI $[2.08, 2.40]$, $p < 0.001$. 
The difference in the perceived competence of the agents significantly altered stated pairwise preferences, $\mathrm{OR} = 0.78$, 95\% CI $[0.73, 0.84]$, $p < 0.001$. 
The effect of the joint intercept was not significant, $p = 0.23$.
The model achieves an AIC of $1610.2$ and $R^2_m = 0.435$.

\begin{figure}[h]
	\centering
    \hspace*{-5em}\includegraphics[height=4.857cm]{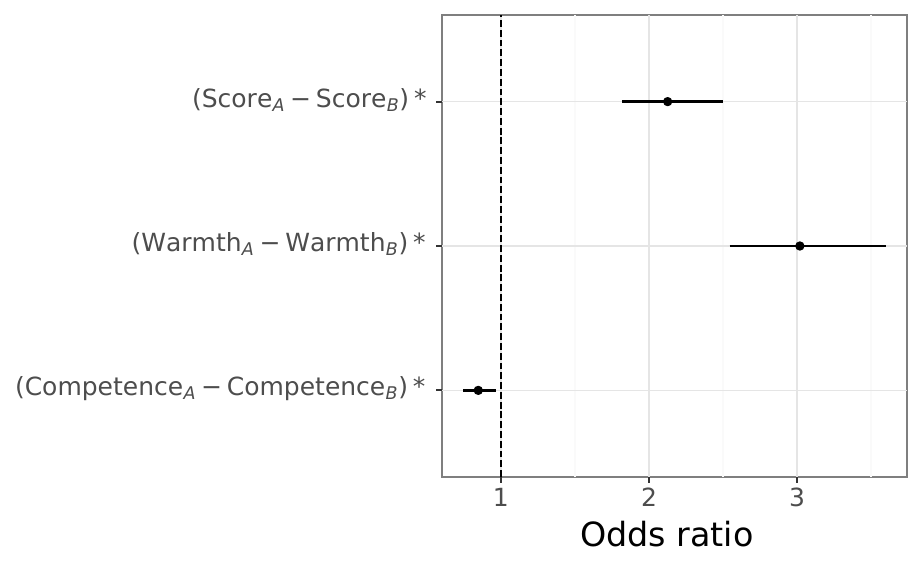}
    \captionsetup{width=.75\textwidth}
	\caption{Odds ratios in Study 1. Asterisks indicate that predictors are centered and normalized by one standard deviation to permit fair comparison between their effect sizes.}
	\label{fig:app/study_1_odds_ratios}
\end{figure}

A final GLMM directly compared the predictiveness of social perceptions against that of the score that participants received. For this model, we standardized the predictors, allowing direct comparison between the model coefficients. The GLMM predicted pairwise preferences from: a joint intercept; standardized variables for perceived warmth, perceived competence, and score; and a random intercept for participant.
All three fixed effects were significant within this model (Figure~\ref{fig:app/study_1_odds_ratios}). One standard deviation of change in the warmth variable (difference in perceived warmth between two agents) had the largest standardized effect on pairwise preferences, $\mathrm{OR} = 3.02$, 95\% CI $[2.54, 3.60]$, $p < 0.001$. One standard deviation in score had the next largest effect, $\mathrm{OR} = 2.13$, 95\% CI $[1.82, 2.50]$, $p < 0.001$. One standard deviation in the competence variable (difference in perceived competence between two agents) had the effect with the smallest magnitude, $\mathrm{OR} = 0.85$, 95\% CI $[0.75, 0.96]$, $p = 0.012$.
The effect of the joint intercept was significant, $\mathrm{OR} = 1.18$, 95\% CI $[1.03, 1.36]$, $p = 0.012$. 

Finally, a linear model predicted post-game impression sentiment from an intercept, perceived warmth, and perceived competence.
Perceived warmth significantly correlated with impression sentiment, $\beta = 0.13$, 95\% CI $[0.09, 0.16]$, $p < 0.001$. Perceived competence did not exhibit a significant relationship with sentiment, $p = 0.24$.
The effect of the intercept was not significant, $p = 0.52$. 

\subsubsection{Study 2}

{As an initial test, we sought to validate the effect of shifting the incentive structure on participants' rewards. A linear mixed-effects model predicted participant reward from a categorical variable representing the study number, with Study 1 as the reference level. As expected, the study variable exerted a significant effect on participant reward, $\beta = 27.3$, 95\% CI $[26.5, 28.0]$, $p < 0.001$.}

As in Study 1, we expect that the social perception questions will exhibit suitable psychometric properties. Table~\ref{tab:study_2_consistency} shows the high consistency with which participants evaluated a given agent on a given trait. A mixed ANOVA showed that social perceptions varied more between traits than within a particular trait, $F_{4,4650} = 88.5$, $p < 0.001$. Finally, both the composite warmth measure ($\rho = 0.92$) and the composite competence measure ($\rho = 0.91$) exhibit high internal consistency, as measured by the Spearman-Brown formula.

\begin{figure}[h]
	\centering
    \subfloat[SVO.]{\includegraphics[height=3.518cm]{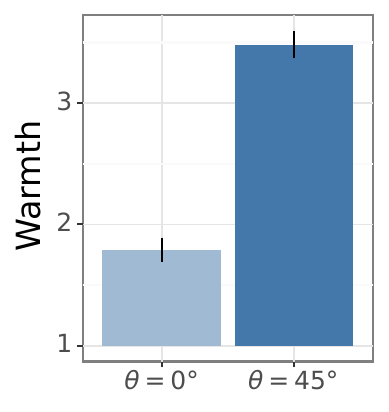} \label{fig:app/study_2_main_effects/a}} \hspace{1em}
    \subfloat[Trembling hand.]{\includegraphics[height=3.518cm]{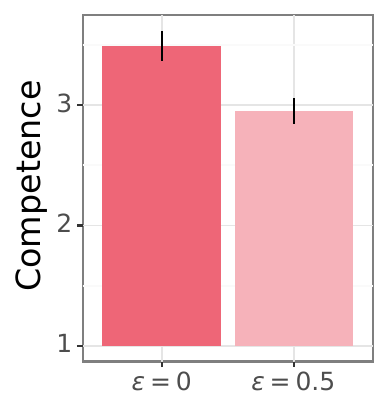} \label{fig:app/study_2_main_effects/b}}
    \captionsetup{width=.75\textwidth}
	\caption{Main effects of the SVO and trembling-hand components on social perceptions in Study 2. (a) An agent's SVO parameter significantly affected warmth perceptions, $p < 0.001$. (b) The trembling-hand component significantly influenced competence evaluations, $p < 0.001$. Error bars reflect 95\% confidence intervals.}
	\label{fig:app/study_2_main_effects}
\end{figure}

\begin{figure}[h]
	\centering
    \subfloat[Warmth.]{\includegraphics[height=4.5cm]{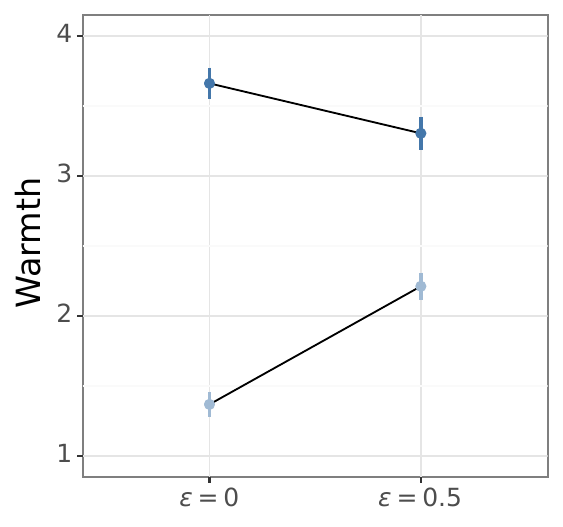} \label{fig:app/study_2_interaction_effects/a}} \hspace{1em}
    \subfloat[Competence.]{\includegraphics[height=4.5cm]{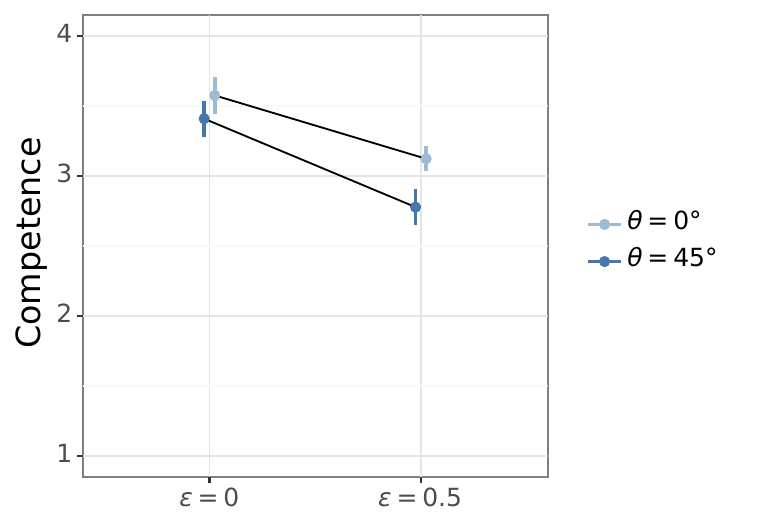} \label{fig:app/study_2_interaction_effects/b}}
    \captionsetup{width=.75\textwidth}
	\caption{Interaction between the SVO and trembling-hand components on social perceptions in Study 2. Error bars reflect 95\% confidence intervals.}
	\label{fig:app/study_2_interaction_effects}
\end{figure}

A two-way mixed ANOVA modeled the effects of $\theta$, $\epsilon$, and their interaction on warmth perceptions. Figure~\ref{fig:app/study_2_main_effects/a} depicts the main effect of $\theta$ on warmth judgments (that is, marginalized over $\epsilon$ values), and Figure~\ref{fig:app/study_2_interaction_effects/a} visualizes the full two-way interaction. The following means and standard deviations help describe main effects by marginalizing over all other variables. Participants assessed $\theta = 45\degree$ agents ($m = 3.48$, $sd = 1.18$) as significantly warmer than $\theta = 0\degree$ agents ($m = 1.79$, $sd = 1.00$), $F_{1,1086} = 981.9$, $p < 0.001$. The $\epsilon$ parameter also exerted a significant effect on warmth evaluations, $F_{1,1086} = 20.3$, $p = 0.002$. Participants perceived $\epsilon = 0.5$ agents ($m = 2.76$, $sd = 1.23$) as significantly warmer than $\epsilon = 0$ agents ($m = 2.52$, $sd = 1.51$). Finally, the effect of the interaction between the $\theta$ and $\epsilon$ parameters was significant, $F_{1,1086} = 123.4$, $p < 0.001$.

A two-way mixed ANOVA modeled the effects of $\theta$, $\epsilon$, and their interaction on competence evaluations. Figure~\ref{fig:app/study_2_main_effects/b} depicts the main effect of $\epsilon$ on competence evaluations (that is, marginalized over $\theta$ values), and Figure~\ref{fig:app/study_2_interaction_effects/b} visualizes the full two-way interaction. The following means and standard deviations help describe main effects by marginalizing over all other variables. Participants judged $\epsilon = 0$ agents ($m = 3.49$, $sd = 1.27$) as significantly more competent than $\epsilon = 0.5$ agents ($m = 2.95$, $sd = 1.09$), $F_{1,1086} = 76.0$, $p < 0.001$. The $\theta$ parameter also exerted a significant effect on competence evaluations, with $\theta = 0\degree$ agents ($m = 3.35$, $sd = 1.22$) perceived as significantly more competent than $\theta = 45\degree$ agents ($m = 3.09$, $sd = 1.20$), $F_{1,1086} = 16.9$, $p < 0.001$. Finally, the effect of the interaction between the $\theta$ and $\epsilon$ parameters was not significant, $F_{1,1086} = 2.1$, $p = 0.15$.

A sequence of fractional response models assessed the predictive value of the hypothesized predictors for participants' stated pairwise preferences. We construct each model through a GLMM with a binomial link function, normalizing the full preference scale with zero and one as its endpoints. To ensure model comparability, all fractional response models predict the exact same outcome variable.

\begin{figure}[h]
	\centering
    \includegraphics[height=4.75cm]{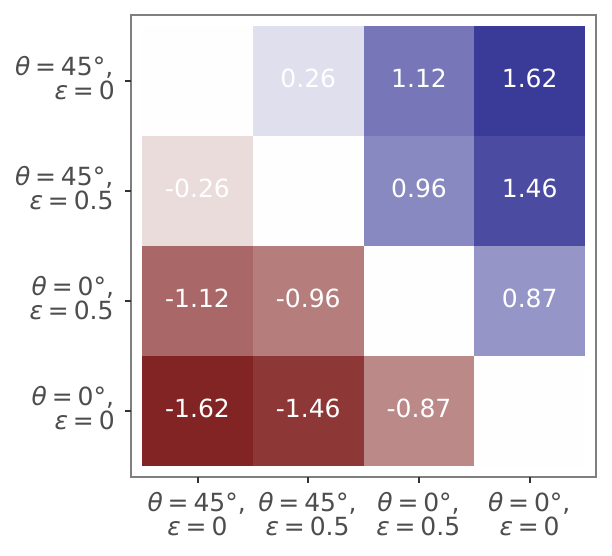}
    \captionsetup{width=.75\textwidth}
	\caption{Pairwise preferences for different agents in Study 2. Cell values represent self-reported preferences for the row agent over the column agent.}
	\label{fig:app/study_2_preference_matrix}
\end{figure}

The first GLMM predicted stated preferences from a joint intercept, a fixed effect for the identities of the agents being compared, and a random effect for the participant.
The underlying identities of the agents significantly affected the participant's stated pairwise preferences, $\chi^2(11) = 549.7$, $p < 0.001$. Figure~\ref{fig:app/study_2_preference_matrix} shows mean pairwise preferences calculated for each pair of agent identities. 
The effect of the joint intercept was not significant, $p = 0.15$.
The model achieves an AIC of $1608.7$ and $R^2_m = 0.403$.

The second GLMM predicted stated preferences from a joint intercept, a fixed effect for the difference in scores the participant received when playing with each of the agents being compared, and a random effect for the participant.
The difference in scores earned by the participant significantly influenced stated pairwise preferences, $\mathrm{OR} = 1.06$, 95\% CI $[1.06, 1.07]$, $p < 0.001$. 
The effect of the joint intercept was not significant, $p = 0.81$.
The model achieves an AIC of $2049.4$ and $R^2_m = 0.214$.

The third GLMM predicted stated preferences from a joint intercept, a fixed effect for the difference in perceived warmth for each of the agents being compared, a fixed effect for the difference in perceived competence for each of the agents, and a random effect for the participant.
The difference in the perceived warmth of the agents significantly altered stated pairwise preferences, $\mathrm{OR} = 2.63$, 95\% CI $[2.42, 2.89]$, $p < 0.001$. 
The difference in the perceived competence of the agents significantly affected stated pairwise preferences, $\mathrm{OR} = 0.82$, 95\% CI $[0.76, 0.88]$, $p < 0.001$. 
The effect of the joint intercept was not significant, $p = 0.99$.
The model achieves an AIC of $1510.4$ and $R^2_m = 0.496$.

\begin{figure}[h]
	\centering
    \hspace*{-5em}\includegraphics[height=4.857cm]{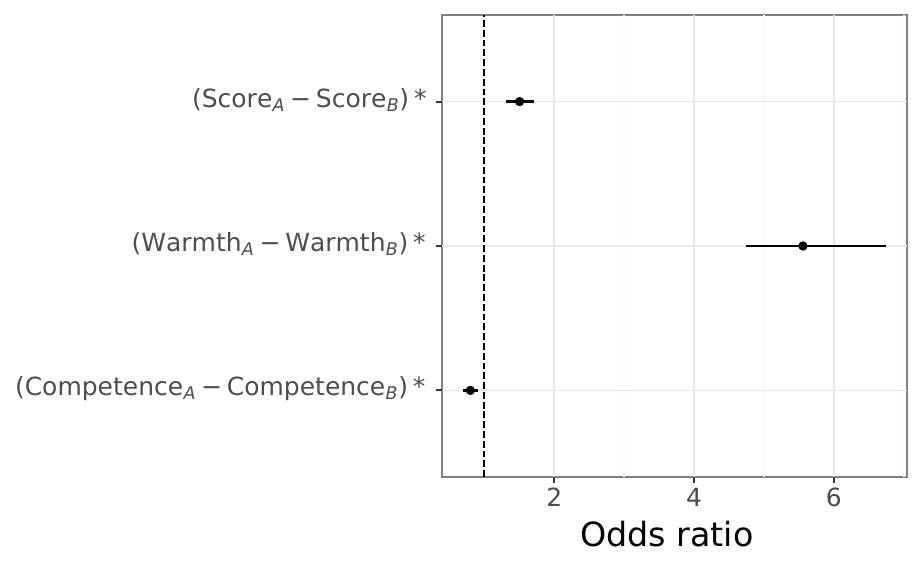}
    \captionsetup{width=.75\textwidth}
	\caption{Odds ratios in Study 2. Asterisks indicate that predictors are centered and normalized by one standard deviation to permit fair comparison between their effect sizes.}
	\label{fig:app/study_2_odds_ratios}
\end{figure}

A final GLMM directly compared the predictiveness of social perceptions against that of the score that participants received. For this model, we standardized the predictors, allowing direct comparison between the model coefficients. The GLMM predicted pairwise preferences from: a joint intercept; standardized variables for perceived warmth, perceived competence, and score; and a random intercept for participant.
All three fixed effects were significant within this model (Figure~\ref{fig:app/study_2_odds_ratios}). One standard deviation of change in the warmth variable (difference in perceived warmth between two agents) had the largest standardized effect on pairwise preferences, $\mathrm{OR} = 5.56$, 95\% CI $[4.74, 6.74]$, $p < 0.001$. One standard deviation in score had the next largest effect, $\mathrm{OR} = 1.50$, 95\% CI $[1.31, 1.71]$, $p < 0.001$. One standard deviation in the competence variable (difference in perceived competence between two agents) had the effect with the smallest magnitude, $\mathrm{OR} = 0.80$, 95\% CI $[0.70, 0.91]$, $p < 0.001$.
The effect of the joint intercept was not significant, $p = 0.49$. 

Finally, a linear model predicted post-game impression sentiment from an intercept, perceived warmth, and perceived competence.
Perceived warmth significantly correlated with impression sentiment, $\beta = 0.12$, 95\% CI $[0.09, 0.15]$, $p < 0.001$ (Figure~\ref{fig:study_2_sentiment/a}). Perceived competence also significantly correlated with sentiment, $\beta = 0.04$, 95\% CI $[0.00, 0.08]$, $p = 0.037$ (Figure~\ref{fig:study_2_sentiment/b}).
The effect of the intercept on impression sentiment was significant, $\beta = -0.15$, 95\% CI $[-0.26, -0.05]$, $p = 0.005$. 

\subsubsection{Study 3}

In Study 3, each participant interacted with a single agent, and thus only provided a single measurement point for warmth judgments and competence judgments. As a result, the ANOVAs and regressions reported below do not incorporate random effects for participants.

An ANOVA indicated that social perceptions varied more between traits than within a particular trait, $F_{3,1200} = 54.9$, $p < 0.001$. Both the composite warmth measure ($\rho = 0.85$) and the composite competence measure ($\rho = 0.86$) exhibit high internal consistency, as measured by the Spearman-Brown formula.

\begin{figure}[h]
	\centering
    \subfloat[SVO.]{\includegraphics[height=3.518cm]{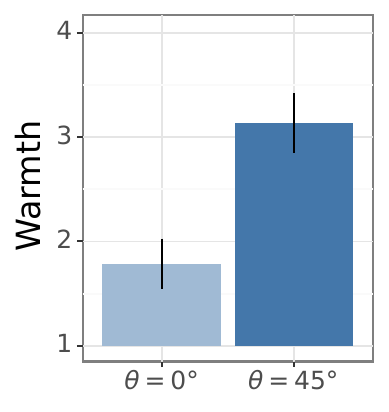} \label{fig:app/study_3_main_effects/a}} \hspace{1em}
    \subfloat[Trembling hand.]{\includegraphics[height=3.518cm]{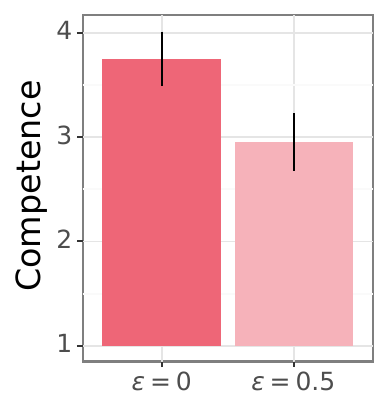} \label{fig:app/study_3_main_effects/b}}
    \captionsetup{width=.75\textwidth}
	\caption{Main effects of the SVO and trembling-hand components on social perceptions in Study 3. (a) The SVO component significantly altered warmth judgments, ${p < 0.001}$. (b) Similarly, an agent's trembling-hand parameterization significantly influenced perceived competence, ${p < 0.001}$. Error bars reflect 95\% confidence intervals.}
	\label{fig:app/study_3_main_effects}
\end{figure}

\begin{figure}[h]
	\centering
    \subfloat[Warmth.]{\includegraphics[height=4.5cm]{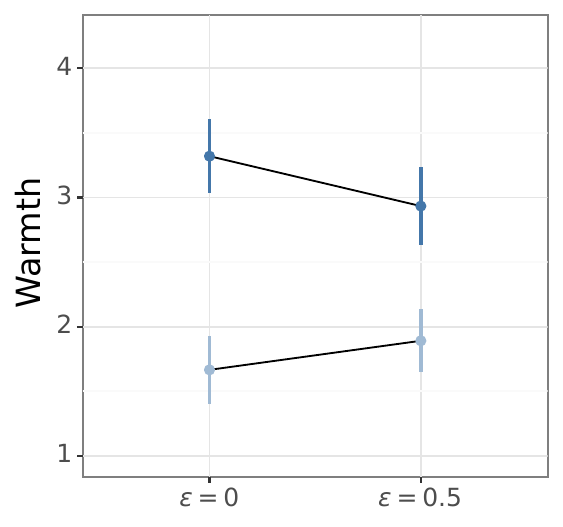} \label{fig:app/study_3_interaction_effects/a}} \hspace{1em}
    \subfloat[Competence.]{\includegraphics[height=4.5cm]{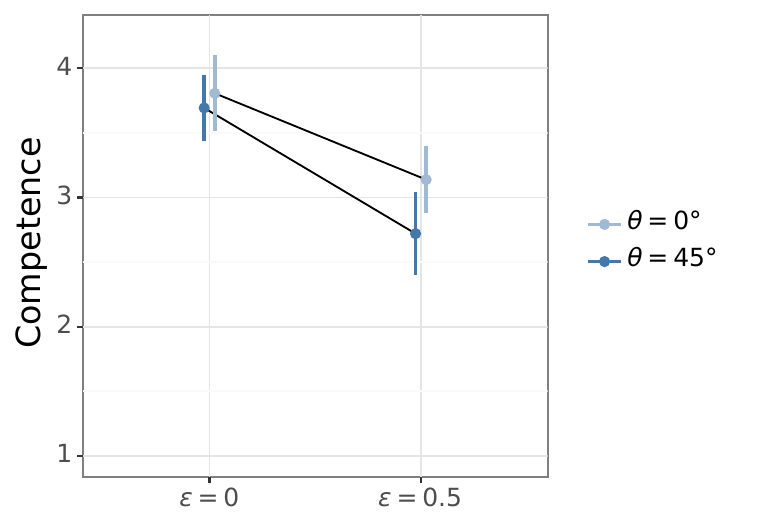} \label{fig:app/study_3_interaction_effects/b}}
    \captionsetup{width=.75\textwidth}
	\caption{Interaction between the SVO and trembling-hand components on social perceptions in Study 3. Error bars reflect 95\% confidence intervals.}
	\label{fig:app/study_3_interaction_effects}
\end{figure}

A two-way ANOVA modeled the effects of $\theta$, $\epsilon$, and their interaction on warmth evaluations. Figure~\ref{fig:app/study_3_main_effects/a} depicts the main effect of $\theta$ on perceived warmth (that is, marginalized over $\epsilon$ values), and Figure~\ref{fig:app/study_3_interaction_effects/a} visualizes the full two-way interaction. The following means and standard deviations help describe main effects by marginalizing over all other variables. Participants perceived $\theta = 45\degree$ agents ($m = 3.14$, $sd = 1.22$) as significantly warmer than $\theta = 0\degree$ agents ($m = 1.79$, $sd = 1.09$), $F_{1,297} = 103.4$, $p < 0.001$. The $\epsilon$ parameter did not exert a significant effect on warmth judgments, $F_{1,297} = 0.2$, $p = 0.62$. Finally, the effect of the interaction between the $\theta$ and $\epsilon$ parameters was significant, $F_{1,297} = 5.3$, $p = 0.022$.

A two-way ANOVA modeled the effects of $\theta$, $\epsilon$, and their interaction on competence judgments. Figure~\ref{fig:app/study_3_main_effects/b} depicts the main effect of $\epsilon$ on competence evaluations (that is, marginalized over $\theta$ values), and Figure~\ref{fig:app/study_3_interaction_effects/b} visualizes the full two-way interaction. The following means and standard deviations help describe main effects by marginalizing over all other variables. Participants judged $\epsilon = 0$ agents ($m = 3.75$, $sd = 1.15$) as significantly more competent than $\epsilon = 0.5$ agents ($m = 2.95$, $sd = 1.23$), $F_{1,297} = 35.3$, $p < 0.001$. The $\theta$ parameter did not exert a significant effect on competence evaluations, $F_{1,297} = 3.8$, $p = 0.05$. Finally, the effect of the interaction between the $\theta$ and $\epsilon$ parameters was not significant, $F_{1,297} = 1.2$, $p = 0.26$.

A sequence of logistic regressions evaluated the predictive value of the hypothesized predictors for participants' revealed preferences.

The first logistic regression predicted revealed preferences from an intercept and the identity of the agent.
The underlying identity of the agent significantly affected the participant's revealed preferences, $\chi^2(3) = 45.4$, $p < 0.001$.
The effect of the intercept on revealed preferences was significant, $\mathrm{OR} = 1.88$, 95\% CI $[1.18, 3.07]$, $p = 0.009$.
The model achieves an AIC of $372.5$ and $R^2 = 0.188$.

The second logistic regression predicted revealed preferences from an intercept and the score earned by the participant when playing with the agent.
The score earned by the participant exerted a significant influence on revealed preferences, $\mathrm{OR} = 1.06$, 95\% CI $[1.03, 1.08]$, $p < 0.001$. 
The effect of the intercept was also significant, $\mathrm{OR} = 0.16$, 95\% CI $[0.08, 0.31]$, $p < 0.001$.
The model achieves an AIC of $390.5$ and $R^2 = 0.101$.

The third logistic regression predicted revealed preferences from an intercept, perceived warmth, and perceived competence.
Perceived warmth significantly predicted revealed preferences, $\mathrm{OR} = 2.10$, 95\% CI $[1.69, 2.65]$, $p < 0.001$. 
Perceived competence did not significantly predict revealed preferences, $p = 0.88$. 
The effect of the intercept was not significant, $p = 0.23$.
The model achieves an AIC of $356.2$ and $R^2 = 0.242$.

\begin{figure}[h]
	\centering
    \hspace*{-3em}\includegraphics[height=4.857cm]{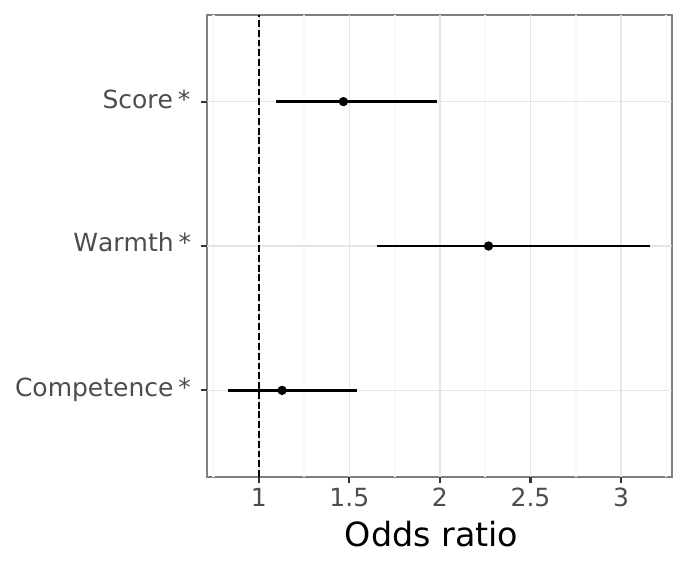}
    \captionsetup{width=.75\textwidth}
	\caption{Odds ratios in Study 3. Asterisks indicate that predictors are centered and normalized by one standard deviation to permit fair comparison between their effect sizes.}
	\label{fig:app/study_3_odds_ratios}
\end{figure}

A final logistic regression directly compared the predictiveness of social perceptions against that of the score that participants received. For this model, we standardized the predictors, allowing direct comparison between the model coefficients. The model predicted revealed preferences from an intercept and standardized variables for perceived warmth, perceived competence, and score (Figure~\ref{fig:app/study_3_odds_ratios}).
The effects of perceived warmth and of score were significant within the model. One standard deviation of change in perceived warmth had the larger effect on revealed preferences, $\mathrm{OR} = 2.27$, 95\% CI $[1.65, 3.16]$, $p < 0.001$. One standard deviation in score had a smaller effect, $\mathrm{OR} = 1.47$, 95\% CI $[1.09, 1.98]$, $p = 0.011$. The effect of perceived competence was not significant, $p = 0.44$.
The effect of the intercept was significant, $\mathrm{OR} = 1.18$, 95\% CI $[1.03, 1.36]$, $p = 0.006$. 

Finally, a linear model predicted post-game impression sentiment from an intercept, perceived warmth, and perceived competence.
Perceived warmth significantly correlated with impression sentiment, $\beta = 0.14$, 95\% CI $[0.10, 0.18]$, $p < 0.001$ (Figure~\ref{fig:study_3_sentiment/a}). Perceived competence also significantly correlated with sentiment, $\beta = 0.04$, 95\% CI $[0.00, 0.08]$, $p = 0.041$ (Figure~\ref{fig:study_3_sentiment/b}).
The effect of the intercept on impression sentiment was significant, $\beta = -0.15$, 95\% CI $[-0.25, -0.05]$, $p = 0.002$. 

\clearpage
\null
\vfill
\vfill

\clearpage
\null
\vfill

\vfill
\vfill

\clearpage
\null
\vfill
\vfill

\clearpage
\null
\vfill
\vfill

\clearpage
\null
\vfill
\vfill

\clearpage
\null
\vfill
\vfill

\vfill

\clearpage
\null
\vfill
\vfill

\clearpage
\null
\vfill
\vfill

\clearpage
\null
\vfill

\vfill

\vfill

\clearpage
\null
\vfill

\vfill

\vfill

\clearpage
\null
\vfill

\vfill

\vfill

\clearpage
\null
\vfill

\vfill

\vfill

\clearpage
\null
\vfill

\vfill

\vfill

\end{document}